\documentclass[sigconf,nonacm]{acmart}

\usepackage{xcolor}

\usepackage{fancyhdr}
\usepackage{adjustbox}
\usepackage{subcaption}

\usepackage{graphicx} 
\usepackage{geometry}
\usepackage{multirow}
\usepackage{subcaption} 
\usepackage{hyperref}

\usepackage{caption}
\usepackage{subcaption}
\usepackage{accsupp}

\usepackage{tabularx}

\usepackage{xfp}
\usepackage{array}
\usepackage{booktabs}
\usepackage{tikz}
\usetikzlibrary{patterns}

\usepackage{placeins}

\usepackage{listings}

\usepackage[T1]{fontenc}
\usepackage{xparse}
\usepackage{enumitem}
\setlist[description]{
  font={\sffamily\bfseries},
  labelsep=25pt,
  labelwidth=\transcriptlen,
  leftmargin=!,
}

\newlength{\transcriptlen}

\NewDocumentCommand {\setspeaker} { mo } {%
  \IfNoValueTF{#2}
  {\expandafter\newcommand\csname#1\endcsname{\item[#1:]}}%
  {\expandafter\newcommand\csname#1\endcsname{\item[#2:]}}%
  \IfNoValueTF{#2}
  {\settowidth{\transcriptlen}{#1}}%
  {\settowidth{\transcriptlen}{#2}}%
}

\definecolor{likert1}{HTML}{c65154}
\definecolor{likert2}{HTML}{e47961}
\definecolor{likert3}{HTML}{fad4ac}
\definecolor{likert4}{HTML}{ffffe0}
\definecolor{likert5}{HTML}{bce2cf}
\definecolor{likert6}{HTML}{579eb9}
\definecolor{likert7}{HTML}{397aa8}

\newcommand{\likertpattern}[1]{%
  \ifcase#1 %
    \or vertical lines
    \or north west lines
    \or grid
    \or crosshatch dots
    \or horizontal lines
    \or north east lines
    \or crosshatch
    \else unknown pattern%
  \fi
}

\newcommand{\likertcolor}[1]{%
  \ifcase#1
    black
    \or likert1
    \or likert2
    \or likert3
    \or likert4
    \or likert5
    \or likert6
    \or likert7
    \else black%
  \fi
}

\usepackage{tikz}
\usetikzlibrary{patterns,calc}

\usepackage[normalem]{ulem}

\usepackage{ccicons}
\usepackage{pdfcomment}
\usepackage{float}
\usepackage{caption}
\usepackage{makecell} 
\usepackage{array}    

\begin{document}
\title[NeuroBridge: Using Generative AI to Bridge Cross-neurotype Communication Differences through Neurotypical Perspective-taking]{NeuroBridge: Using Generative AI to Bridge Cross-neurotype Communication Differences through Neurotypical Perspective-taking}

\setcopyright{none}
\acmConference[ASSETS '25]{The 27th International ACM SIGACCESS Conference
on Computers and Accessibility}{October 26--29, 2025}{Denver, CO, USA}
\acmBooktitle{The 27th International ACM SIGACCESS Conference on Computers
and Accessibility (ASSETS '25), October 26--29, 2025, Denver, CO, USA}
\acmDOI{10.1145/3663547.3746337}
\acmISBN{979-8-4007-0676-9/2025/10}

\author{Rukhshan Haroon}
\affiliation{%
  \department{Computer Science} 
  \institution{Tufts University}
  \state{MA}
  \country{United States}
}
\email{rukhshan.haroon@tufts.edu}

\author{Kyle Wigdor}
\affiliation{%
  \department{Computer Science} 
  \institution{Tufts University}
  \state{MA}
  \country{United States}
}
\email{kyle.wigdor@tufts.edu}

\author{Katie Yang}
\affiliation{%
  \department{Computer Science} 
  \institution{Tufts University}
  \state{MA}
  \country{United States}
}
\email{zihan.yang@tufts.edu}

\author{Nicole Toumanios}
\affiliation{%
  \department{Child Studies and Human Development, Cognitive Science} 
  \institution{Tufts University}
  \state{MA}
  \country{United States}
}
\email{nicole.toumanios@tufts.edu}

\author{Eileen T. Crehan}
\affiliation{%
  \department{Eunice Kennedy Shriver Center}  
  \institution{UMass Chan Medical School}
  \state{MA}
  \country{United States}
}
\email{eileen.crehan@umassmed.edu}

\author{Fahad Dogar}
\affiliation{%
  \department{Computer Science} 
  \institution{Tufts University}
  \state{MA}
  \country{United States}
}
\email{fahad@cs.tufts.edu}


\date{September 2023}

\begin{abstract}

\begin{figure*}[t] 
    \centering
    \includegraphics[width=\textwidth]{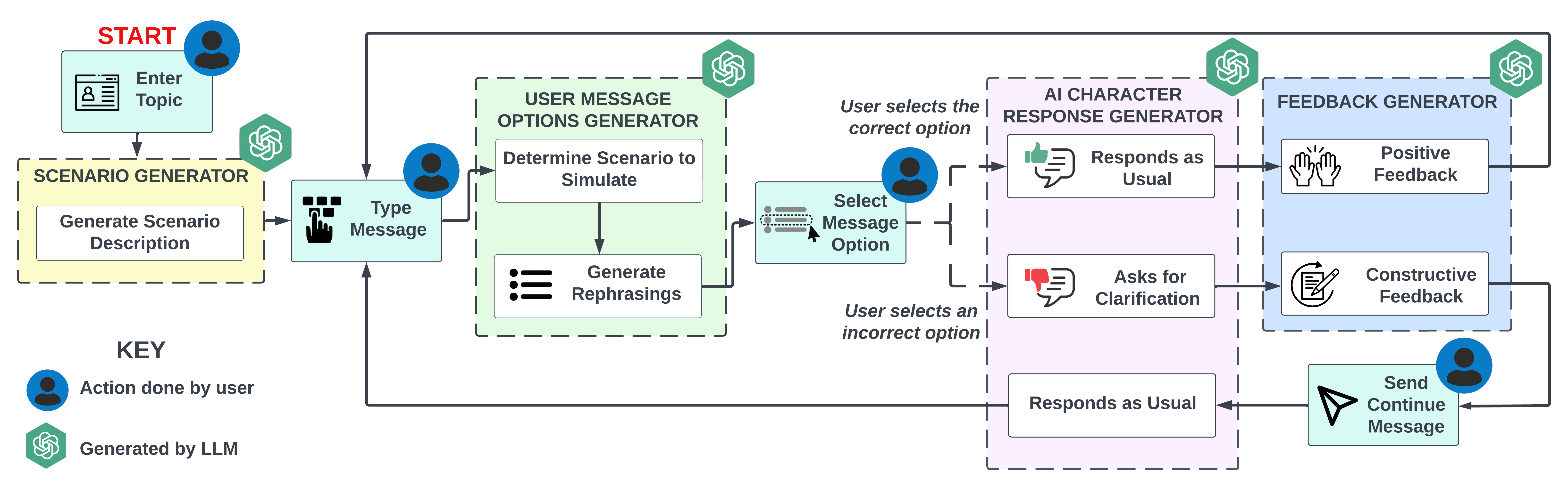} 
    \Description[Flowchart of the NeuroBridge system.]{Flowchart of the NeuroBridge system architecture. Users begin by entering a topic, which is used by the Scenario Generator to create a scenario. The user then types in a message, which is rephrased by the Message Options Generator to create three message options. The user selects one of these options, and based on this choice, the Response Generator produces a response from the AI character. The Feedback Generator then provides positive or constructive feedback based on the user's choice. If constructive feedback was given, the user must send a continue message to clarify, prompting a response from the AI character. The process then repeats as the user types another message.}
    \caption{NeuroBridge architecture and interaction flow. Users begin by entering a topic and then engage in a loop of sending messages, receiving responses, and getting feedback.}
    \label{fig:architecture}
\end{figure*}
Communication challenges between autistic and neurotypical individuals stem from a \textit{mutual} lack of understanding of each other's distinct, and often contrasting, communication styles. Yet, autistic individuals are expected to adapt to neurotypical norms, making interactions inauthentic and mentally exhausting for them. To help redress this imbalance, we build NeuroBridge, an online platform that utilizes large language models (LLMs) to simulate: a) an AI character that is direct and literal, a style common among many autistic individuals, and b) four cross-neurotype communication scenarios in a feedback-driven conversation between this character and a neurotypical user. Through NeuroBridge, neurotypical individuals gain a firsthand look at autistic communication, and reflect on \textit{their} role in shaping cross-neurotype interactions. In a user study with 12 neurotypical participants, we find that NeuroBridge improved their understanding of how autistic people may interpret language differently, with all describing autism as a social difference that “needs understanding by others” after completing the simulation. Participants valued its personalized, interactive format and described AI-generated feedback as "constructive", "logical" and "non-judgmental". Most perceived the portrayal of autism
in the simulation as accurate, suggesting that users may readily accept AI-generated (mis)representations of disabilities. To conclude, we discuss design implications for disability representation in AI, the need for making NeuroBridge more personalized, and LLMs' limitations in modeling complex social scenarios.

\end{abstract}

\maketitle

\noindent\textbf{Author's Note: }This paper has been conditionally accepted for publication in the Proceedings of the ACM SIGACCESS Conference on Computers and Accessibility (ASSETS 2025), to be held October 26–29, 2025, in Denver, CO, USA. This is a preprint version.

\section{Introduction}
Autism Spectrum Disorder (ASD) is a complex neurodevelopmental condition marked by differences in communication, cognition, sensory processing, and social behavior compared to neurotypical development \cite{diagnostic-american, autism-hodges, patterns-grossi}. It is one of the most common neurodevelopmental conditions in the U.S., affecting an estimated 1 in 45 adults \cite{prevalence-c}. Key traits of autistic communication include a preference for a direct conversational style \cite{my-barros, twips-haroon}, literal language \cite{literalism-vicente, figurative-kalandadze}, and minimal use of social cues 
\cite{neural-hirsch, impaired-griffiths}. These often contrast with neurotypical communication norms, which involve phatic exchanges, implied intent, and social nuance \cite{my-barros, new-miller, editorial-sagliano, the-clough}. Prior work shows that cross-neurotype communication breakdowns due to these differences can have severe consequences for autistic individuals, such as social exclusion in both online and physical social spaces \cite{my-barros, twips-haroon, neurotypical-sasson, self-perception-finke}, professional setbacks \cite{life-rebholz, aspergers-dokumen}, and barriers to quality healthcare \cite{respect-nicolaidis, barriers-doherty}.

Prior efforts to bridge this divide include technological, educational, and therapy-based interventions \cite{twips-haroon, educational-klefbeck, systematic-chang, communication-wolf}. However, these have predominantly targeted autistic individuals, often pressuring them to conform to neurotypical norms. The double empathy problem underscores that communication challenges between autistic and neurotypical individuals arise from reciprocal misunderstandings, necessitating efforts from both sides to work toward mutual understanding and acceptance \cite{the-milton}. Yet, interventions at the neurotypical end are nearly nonexistent, limited to passive, informational resources that offer no opportunities to practice learned concepts or incentives to get involved \cite{awareness-autistic, autism-kendall}. This imbalance places the burden of adapting communication styles almost entirely on autistic individuals.

As large language model (LLM) powered chatbots like ChatGPT \cite{gpt4-openai} and Character.AI \cite{characterai} gain widespread traction, with tens of millions of users engaging with them daily, LLMs present a powerful, new avenue for designing immersive, scalable, and personalized human-AI interactions. Their ability to generate fluent, human-like text, interpret subtle linguistic cues, and adapt to diverse conversational styles makes them well-suited for simulating real-world communication scenarios, including those involving different neurotypes \cite{generative-park, shanahan-role, designing-cao}. We believe this capability, if utilized responsibly, can be used to engage neurotypical individuals in interactive, personalized learning experiences that cultivate empathy and appreciation for autistic communication styles. While existing applications of LLMs in this space focus on providing communication support to autistic individuals \cite{twips-haroon, jang-its}, we advocate for shared responsibility and shift the focus of intervention to the neurotypical end.


In this paper, we present NeuroBridge, an interactive platform designed to help neurotypical individuals better understand autistic forms of expression, and reflect on how their own behavior shapes cross-neurotype interactions. At its core, NeuroBridge utilizes LLMs to simulate: a) an AI character configured to be direct and literal, a style common among many autistic individuals, and b) four cross-neurotype communication scenarios in a feedback-driven conversation between the character and a neurotypical user. Informed by prior work and vetted by an advisory board of autistic individuals, these scenarios (outlined in Table \ref{tab:scenarios}) reflect common communication challenges faced by autistic individuals \cite{twips-haroon, my-barros, literalism-vicente, figurative-kalandadze}. The character may request clarification from users when needed, and for each scenario, users are given tailored feedback to work through their differences with the character empathically. Through NeuroBridge, neurotypical individuals gain a firsthand look at autistic communication, and learn how to communicate more effectively with autistic individuals. Throughout development, members of the advisory board evaluated NeuroBridge at different stages and provided in-depth design feedback, which was incorporated to ensure the simulation reflected their lived experience as closely as possible.



Through an in-lab user study with 12 neurotypical participants recruited from a university setting, aged 18 to 34, we gather survey and in-depth qualitative data on the perceived usefulness of NeuroBridge, how it shaped participants’ perceptions of autism, their attitudes toward AI feedback, and LLMs' ability to model complex communication scenarios. We find that NeuroBridge improved participants’ understanding of how autistic people may interpret language differently, with all describing autism as a social difference that “needs understanding by others” after completing the simulation. Participants valued the simulation’s personalized, interactive format and described AI-generated feedback as “constructive,” “logical,” and “non-judgmental.” On certain occasions, however, participants found the feedback instructional, which led to feelings of defensiveness. Most perceived the portrayal of autism in the simulation as accurate, suggesting that users may readily accept AI-generated (mis)representations of disabilities. Despite strong overall performance, our findings show that LLMs are more adept at simulating certain social scenarios than others. To conclude our work, we present a discussion around the implications for representing disabilities through AI, the need and opportunities for making NeuroBridge more personalized, and the limitations of LLMs in modeling complex social scenarios. 


To summarize, we make the following key contributions:
\begin{itemize}
    \item Make a case to integrate the double empathy problem's theoretical framework into practice for bridging cross-neurotype communication differences in a neurodiversity-affirming manner.
    
    \item Co-design and implement NeuroBridge with an advisory board of autistic individuals to help neurotypical people better understand autistic forms of expression, and reflect on their role in shaping cross-neurotype interactions.
    
    \item Evaluate NeuroBridge in an IRB-approved study with 12 neurotypical participants, collecting in-depth feedback on the simulation's usefulness and impact on user perceptions of autistic abilities, user attitudes toward AI feedback, and the effect of LLM-driven personalization on user engagement.
    
    \item Present a discussion around design implications for representing disabilities through AI, the need and opportunities for making NeuroBridge more personalized, and the limitations of LLMs in modeling various, complex social scenarios.
\end{itemize}

\newcolumntype{L}[1]{>{\raggedright\arraybackslash}p{#1}}

\begin{table*}[t]
\centering
\renewcommand{\arraystretch}{1.6}
\begin{tabularx}{\textwidth}{X X X X}
\Xhline{2\arrayrulewidth}
\textbf{Scenario / Challenge} & \textbf{Description} & \textbf{Example} & \textbf{Interpretations} \\
\hline
Indirect Speech Act & A statement with an implicit request or intent. & Can you open the window? & A literal question about the possibility of opening the window, or a polite request to open it. \\
\hline
Figurative Expression & A phrase whose meaning goes beyond the literal interpretation of words. & She has a chip on her shoulder. & A literal reference to something on one’s shoulder, or as an idiom, one holds a grudge. \\
\hline
Emoji with Variable Interpretations & An emoji with fluid meaning, dependent on context, tone, and personal experience. & That presentation was on \includegraphics[height=1em, alt={fire emoji}]{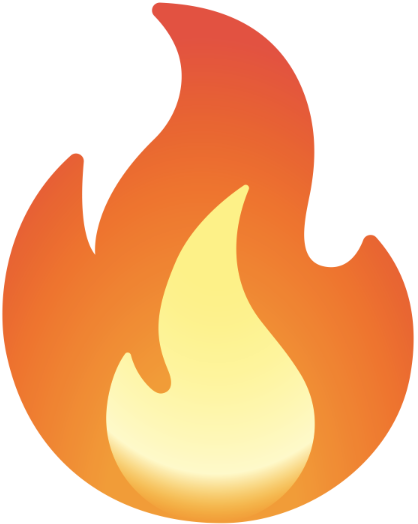} man... & The presentation was impressive, or as sarcasm, it was poor. \\
\hline
Being Misperceived as Blunt & A direct statement by an autistic person that unintentionally comes off as rude or blunt. & I don’t like your idea at all. & A neutral expression of opinion, or harshly expressed criticism. \\
\Xhline{2\arrayrulewidth}
\end{tabularx}
\Description[Table summarizing communication challenges simulated in NeuroBridge.]{Table summarizing communication challenges simulated in NeuroBridge. The first challenge is Indirect Speech Act, described as a statement with an implicit request or intent. The example is "Can you open the window?" which could be interpreted as a literal question about the possibility of opening the window, or a polite request to open it. The second challenge is Figurative Expression, described as a phrase whose meaning goes beyond the literal interpretation of words. The example is "She has a chip on her shoulder," which could be taken as a literal reference to something on one's shoulder, or as an idiom, one holds a grudge. The third challenge is Emoji with Variable Interpretations, described as an emoji with fluid meaning, dependent on context, tone, and personal experience. The example is "That presentation was on [fire emoji] man..." which could be interpreted as either praise or sarcasm, indicating it was poor. The fourth challenge is Being Misperceived Blunt, described as a direct statement by an autistic person that unintentionally comes off as rude or blunt. The example is "I don’t like your idea at all," which may be seen as either a straightforward expression of opinion or as harshly expressed criticism.}
\vspace{0.5em}
\caption{List of communication challenges simulated in NeuroBridge, along with a description, example statement, and the different interpretations of the example that could cause misunderstanding in each scenario.}
\label{tab:scenarios}
\end{table*}

\section{Related Work}

In this section, we review common characteristics of autistic communication, different types of interventions in autism, and the role of technology, particularly LLMs, in advancing them.

\subsection{Characteristics of Autistic Communication}


Numerous studies in disabilities and linguistics research show that key traits of autistic communication include a preference for a direct conversational style \cite{my-barros, twips-haroon}, literal language \cite{literalism-vicente, figurative-kalandadze}, and minimal use of social cues \cite{neural-hirsch, impaired-griffiths}. These autistic norms are known to be rooted in Gricean maxims, which are unwritten rules that guide conversational cooperation by encouraging speakers to be truthful, clear, relevant, and concise \cite{pragmatic-reichow}. For example, when asked, “Can you open the window?”, an autistic individual might respond with a literal “Yes,” interpreting it as a question about ability rather than a request. Similarly, it has been observed that autistic individuals may take figurative expressions such as sarcasm, metaphors, or sexual innuendos at face value \cite{figurative-lampri}. Such literal interpretations can make it hard to infer others’ intentions or navigate the implicit nature of everyday conversation \cite{pragmatic-reichow}. In addition, there is a common misconception that autistic people lack empathy, because their preference for directness may not align with socially accepted norms, and as a result, perceived as bluntness \cite{autistic-kimber}. While these styles are common among many autistic individuals, it is important to note that autism is a spectrum, and they do not apply to all autistic individuals \cite{autism-hodges}.

Communication breakdowns caused by these differences can lead to adverse consequences for autistic individuals, such as social exclusion in online and physical social spaces \cite{my-barros, twips-haroon, neurotypical-sasson, self-perception-finke}, professional setbacks \cite{life-rebholz, aspergers-dokumen}, and barriers to quality healthcare \cite{respect-nicolaidis, barriers-doherty}. For example, autistic users have reported struggling to navigate innuendos in conversations with potential dates on dating applications, and facing harsh reactions on online public forums for being perceived as rude, as opposed to direct and factual, by others \cite{twips-haroon, my-barros}. Similarly, doctors may find it difficult to fully understand an autistic patient’s symptoms if they don’t express themselves in a way that aligns with their expectations \cite{respect-nicolaidis}; in workplace environments, where traits such as diplomacy and politeness are valued, being overly direct can impact relationships with colleagues and slow down career advancement \cite{autism-szechy, an-nicholas}. Therefore, bridging these differences is crucial to improving the day-to-day lives of autistic individuals.

\subsection{Interventions in Autism}

A number of educational, therapeutic, and technological interventions have been developed to support social skills development in autistic individuals. For example, peer-mediated interventions involve typically developing peers to support social interaction and communication development in classroom settings \cite{zhang-effectiveness,rodriguez-peer}. Applied Behavior Analysis (ABA), though controversial in some communities, is commonly used to teach social skills through reinforcement \cite{eckes-comprehensive, choi-patient}. Additionally, research in Human-Computer Interaction (HCI) has advanced support through multiple technology-driven interventions \cite{washington-a, duarte-welcoming, jeong-lexical, washington-superpowerglass, tartaro-authorable, park-aedle, boyd-vrsocial, ringland-dancecraft}. For example, Park et al. combined augmented reality (AR) with drama therapy to facilitate accessible and adaptable language therapy for autistic children \cite{park-aedle}, while Ringland et al. built a whole-body interface to augment dance therapy for autistic children with sensory sensitivities \cite{ringland-dancecraft}. Prior studies highlight the benefits of incorporating technology into interventions, such as greater user engagement \cite{tarantino-on}, access to support \cite{washington-superpowerglass}, and customization for catering individual needs \cite{ahsen-designing}. Broadly, these approaches align with the interventionist and medical models of disability, which view disability as an impairment to be managed or treated through targeted support \cite{birsenden-independent, marks-models}.

On the other hand, the social model of disability emphasizes that disabilities arise not from individual deficits, but from the mismatch between individuals and their social environments \cite{oliver-the, shakespeare-the}. In the specific context of autism, the `double empathy problem' is a concept grounded in neurodiversity theory which posits that communication breakdowns between autistic and non-autistic individuals are bidirectional, stemming from differences in conversational norms and emotional expression \cite{milton-the}; thus, they are the result of mutual misunderstandings, rather than lack of empathy or deficits on part of autistic individuals. As such, interventions should support bidirectional accommodations, rather than focusing solely on training autistic individuals to conform to neurotypical norms \cite{edey-interaction, kimber-autistic}. Yet, interventions at the neurotypical end remain scarce, often limited to passive, informational resources that provide little opportunity for practicing learned concepts or incentive to get involved \cite{imex-autism, awareness-autistic, autism-kendall}. Notable efforts in this space include Autismity -- The Autism Simulator \cite{autismity-about}, a VR-based awareness tool, and The Autism Reality Experience \cite{training-autism}, a mobile sensory van. However, these initiatives are costly, difficult to scale, and primarily focus on the physical and sensory experiences of autistic people. While disability simulation has been critiqued for reinforcing stereotypes and erasing the everyday, systemic experiences of disabled people in the past \cite{tigwell-nuanced, kiger-disability}, recent scholarship suggests that framing it as an educational tool through the social model can improve its effectiveness \cite{barney-disability}. Our work builds on this line of work by engaging participants as active learners rather than passive observers, providing personalized, targeted feedback to encourage accountability, self-reflection, and change, while remaining scalable and cost-effective.

\subsection{LLMs, Communication, and Accessibility}


Recent advances in generative AI have led to the emergence of large language models (LLMs) such as GPT-4 \cite{gpt4-openai} and LLaMa \cite{hugo-llama}. LLMs are capable of generating fluent, human-like text, interpreting subtle linguistic cues, and adapting to a variety of conversational styles \cite{shanahan-role, generative-park}. These capabilities have opened up new possibilities for designing communication support tools for people with diverse needs, including those who are dyslexic, hard-of-hearing, and use augmentative and alternative communication devices \cite{goodman-lampost, alonzo-design, song-emobridge, li-exploring, valencia-the, jang-its, twips-haroon}. Specifically in the context of autism, Jang et al. examined the use of LLMs for communication assistance at the workplace, finding that autistic individuals prefer LLMs over human colleagues due to greater convenience/availability, neutrality, and privacy \cite{jang-its}. Haroon et al. integrated LLMs into an instant messaging application to provide autistic users with in-situ communication assistance, and found that LLMs offer a convenient way for them to seek clarifications, provide a better alternative to tone indicators, and facilitate constructive reflection on writing technique and style \cite{twips-haroon}. Similarly, Barros et al. conducted participatory workshops with autistic social media users to identify their design needs and develop new features to address them; LLMs show promise to power many of the envisioned features \cite{my-barros}. 

However, most of these approaches reinforce a deficit-oriented model of disability by promoting adaptation to dominant social norms. In contrast, our work aims to use LLMs to help neurotypical individuals understand autistic forms of expression, and how their own behaviors shape cross-neurotype communication. Our approach directly aligns with Boyd’s concept of celebratory technologies, which highlights the value of neurodivergent ways of being and advocates for interventions that promote dignity, agency, and social inclusion, rather than focusing on remediation \cite{boyd-conceptualizing}. Researchers have also worked on identifying and mitigating risks, biases, and ethical concerns related to LLMs and disability \cite{park-as, gadiraju-i, adnin-i, glazko-identifying}. Beyond accessibility, LLMs have been used to enhance efficiency, safety, and decision-making across several, critical domains \cite{hedderich-a-piece, codeaid-kazemitabaar-majeed, muhammad-the-poorest,multimodal-yildirim-nur}.

\section{Overview of NeuroBridge}

\begin{figure}[t]  
    \centering
    \fcolorbox{gray!60}{white}{\includegraphics[width=.49\textwidth]{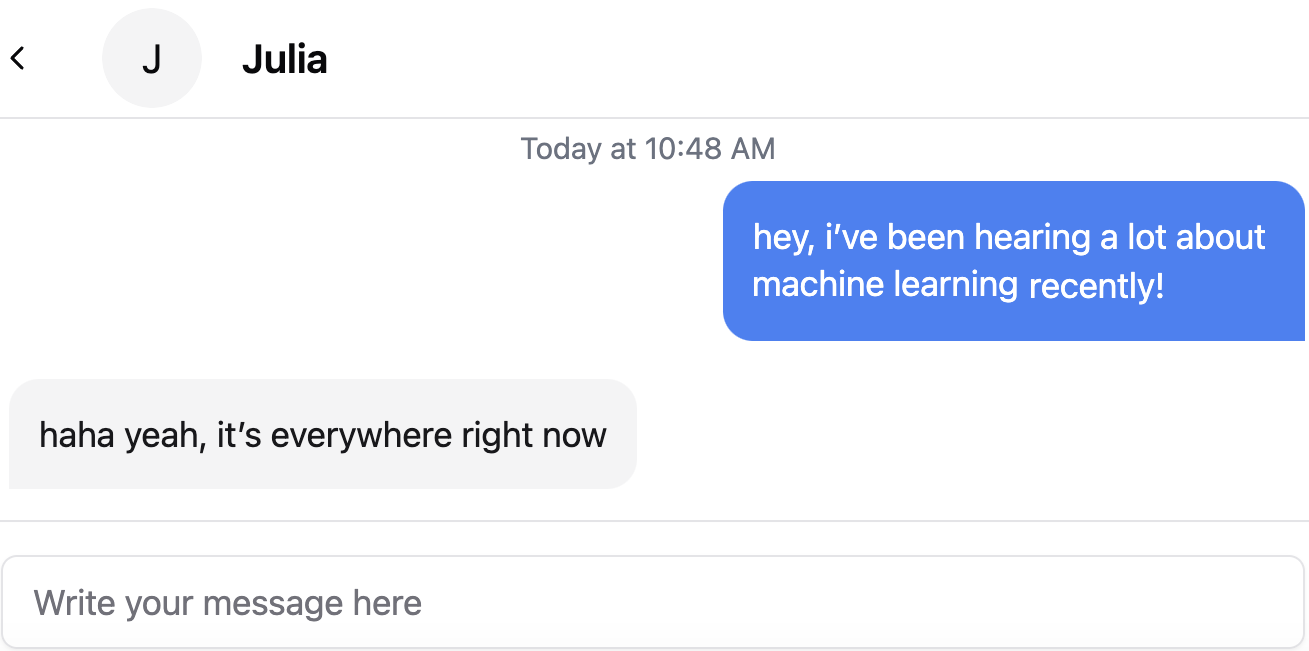}}  
    \caption{The main interface of NeuroBridge is designed to replicate regular messaging apps, making it feel familiar to users. The message in the blue bubble was sent by the user, while the message in the gray bubble was sent by Julia, the AI character.}
    \Description[Screenshot of the main interface of NeuroBridge.]{Screenshot of the main interface of NeuroBridge shows that the user is currently conversing with Julia. In the main panel, a blue message bubble shows the user first sent "hey, i've been hearing a lot about machine learning recently!" followed by a gray message bubble from Julia responding with "haha yeah, it's everywhere right now" Below the conversation history, an input box prompts the user, "Write your message here."}
    \label{fig:overall}
\end{figure}

In this section, we outline the key components, design and implementation of NeuroBridge. 

\begin{figure*}[t]
    \centering
    \begin{subfigure}{0.49\textwidth}
        \centering
        \fcolorbox{gray!60}{white}{
            \begin{minipage}[c][6cm][c]{\textwidth}  
                \centering
                \includegraphics[width=.93\textwidth]{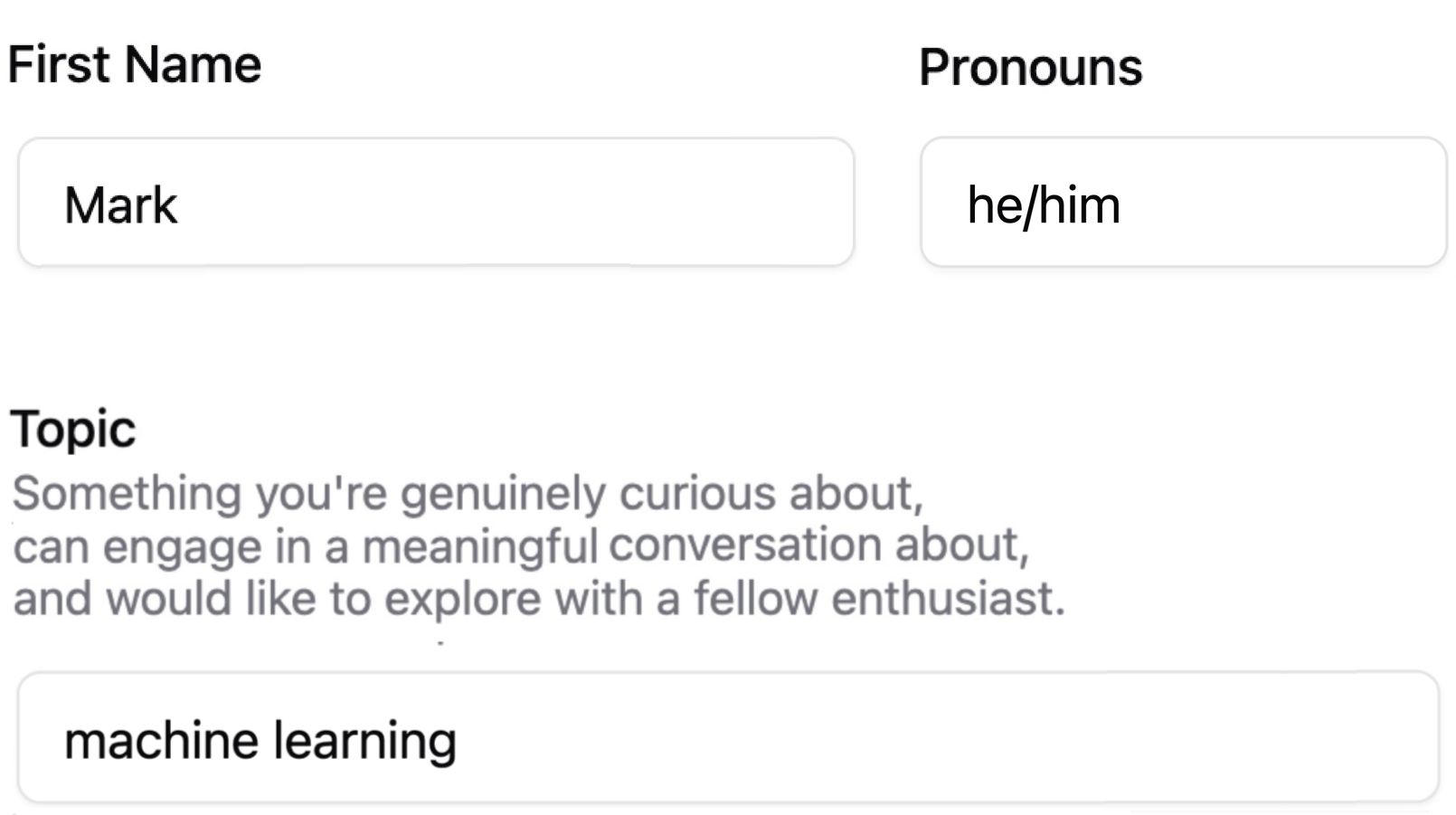}
            \end{minipage}
        }
        \Description[Screenshot of the user registration screen.]{Screenshot of the user registration screen with input fields for first name and pronouns on the first line, a topic of interest field on the second line, and a "Complete Registration" button at the bottom. The user inputs "Mark" as their first name, "he/him" as their pronouns, and "machine learning" as a topic they are interested in.}
        \caption{User registration screen.}
        \label{fig:reg}
    \end{subfigure}
    \hspace{0.01\textwidth}
    \begin{subfigure}{0.49\textwidth}
        \centering
        \fcolorbox{gray!60}{white}{
            \begin{minipage}[c][6cm][c]{\textwidth}  
                \centering
                \includegraphics[width=.9\textwidth]{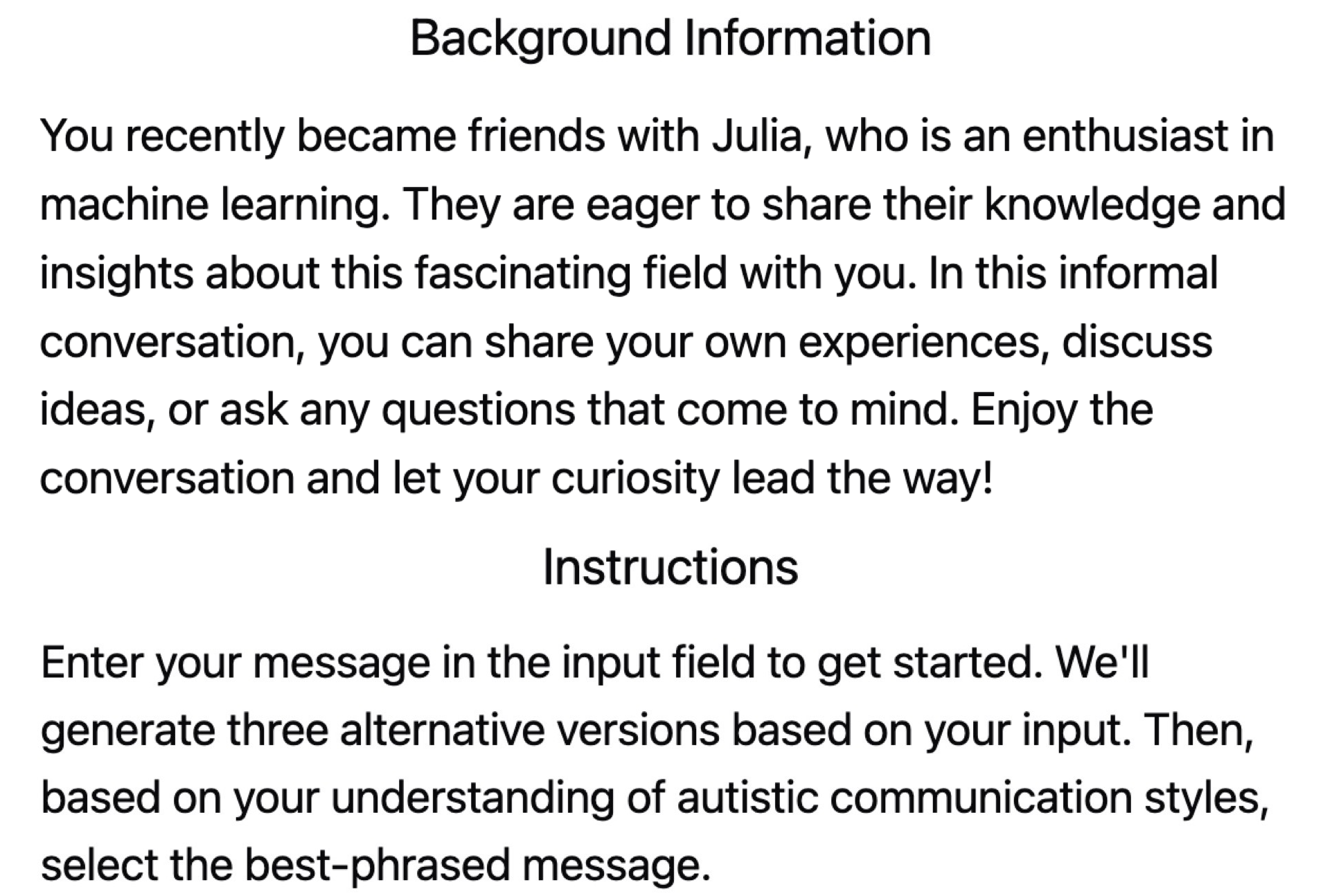}
            \end{minipage}
        }
        \caption{Scenario description screen.}
        \Description[Screenshot of the scenario description screen.]{Screenshot of the scenario description screen with a "Background Information" section that reads, "You recently became friends with Julia, who is an enthusiast in machine learning. They are eager to share their knowledge and insights about this fascinating field with you. In this informal conversation, you can share your own experiences, discuss ideas, or ask any questions that come to mind. Enjoy the conversation and let your curiosity lead the way!"  Below that, an "Instruction" section states, "Enter your message in the input field to get started. We'll generate three alternative versions based on your input. Then, based on your understanding of autistic communication styles, select the best-phrased message." A "Start Chat" button follows. }
        \label{fig:scenario}
    \end{subfigure}
    \caption{The user registration screen, shown in (a), first gathers information about the user. Based on this information, NeuroBridge generates an AI character and a social scenario for the upcoming conversation, as shown in the scenario description screen in (b).}
    \label{fig:reg-sce}
\end{figure*}

\subsection{Components of NeuroBridge}
All components of NeuroBridge (described below) are powered by an LLM. Figure \ref{fig:architecture} captures how these components interact with one another. For each component, we provide the LLM with a unique `prompt' -- a carefully crafted instruction given as input to guide the model's output. A detailed description of our prompting strategy is provided in Section ~\ref{sec:prompting-strategy}. As shown in Figure \ref{fig:overall}, the main interface of NeuroBridge resembles a standard chatting application.

\vspace{1em}
\noindent\textbf{Scenario Generator. } The Scenario Generator creates a conversation scenario tailored for each user based on information they provide about themselves. Figures \ref{fig:reg} and \ref{fig:scenario} show the interface for collecting this information, and an example scenario, respectively. The goal is to center the conversation with the AI character around a topic that is both interesting and relatable for the participant.

\vspace{1em}
\noindent\textbf{Message Options Generator.} The Message Options Generator takes in a user message, and creates three different versions of it, which we call `message options'. This process is shown in Figures \ref{fig:user-input} and \ref{fig:rephrasings}. The message options are similar in meaning to the user's initial message but vary in tone, clarity, or phrasing based on the given scenario (the scenarios are listed in Table \ref{tab:scenarios}). For instance, in the scenario involving indirect speech acts, one option may ask a question directly (“What methods...”), while others phrase the same question ambiguously (“Is there a way...”), as exemplified in Figure \ref{fig:rephrasings}. The user can then select and send one of the three message options. Similarly, for scenarios involving figurative expressions and emojis with variable interpretations, one message option uses literal language or a straightforward emoji, while others express the same idea figuratively. In the scenario involving misperceived bluntness, two options suggest the user found the character’s message blunt, while the third is a neutral response that shows understanding and acceptance of the character’s direct style. Interaction flows for these three scenarios are provided in Appendix \ref{appendix:simulation-flows}. In this way, the message options allow us to trigger different scenarios, while having the user craft the initial message allows for personalizing the simulation experience. 

\begin{figure*}[t]
    \centering
    \fcolorbox{gray!60}{white}{\includegraphics[width=.9\textwidth]{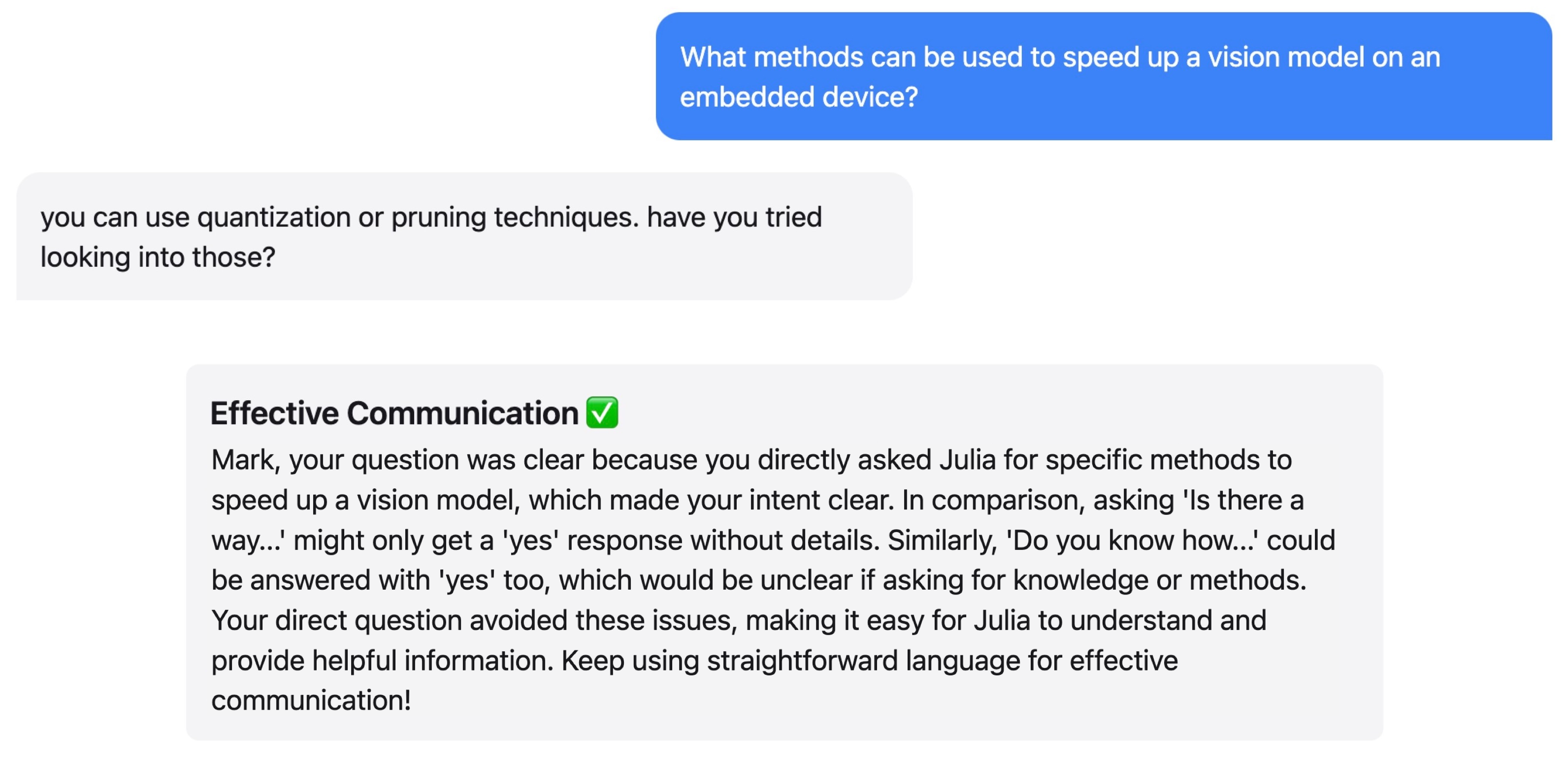}}
    \Description[Screenshot shows the feedback after the most appropriate option is selected.]{Screenshot shows the feedback after the most appropriate option is selected. A blue message bubble first shows the user sent, "What methods can be used to speed up a vision model on an embedded device?" Below, a gray message bubble shows the AI character replied with, "you can use quantization or pruning techniques. have you tried looking into those?" Under this, a feedback panel is titled, "Effective Communication." This section reads, "Mark, your question was clear because you directly asked Julia for specific methods to speed up a vision model, which made your intent clear. In comparison, asking 'Is there a way...' might only get a 'yes' response without details. Similarly, 'Do you know how...' could be answered with 'yes' too, which would be unclear if asking for knowledge or methods. Your direct question avoided these issues, making it easy for Julia to understand and provide helpful information. Keep using straightforward language for effective communication!"}
    \caption{The message option selected by the user is shown in the blue message bubble. In the gray message bubble, the AI character responds as usual because the user selected the correct option. Positive feedback (shown in the bottom gray panel) is provided to reinforce the user's choice and explains why the incorrect options may have caused confusion.}
    \label{fig:corr-feedback}
\end{figure*}

\vspace{1em}
\noindent\textbf{Response Generator.} The Response Generator generates all messages sent by the character. If the user message is unclear, the character's response is a request for clarification. This is shown in Figure \ref{fig:inc-feedback}. If the message is clear, the conversation is continued as usual. This is shown in Figure \ref{fig:corr-feedback}. In the scenario involving misperceived bluntness, if the user indicates that the character's response was blunt, the character follows up in the subsequent message to explain that it was not meant that way.

\vspace{1em}
\noindent\textbf{Feedback Generator.} The Feedback Generator generates scenario-specific feedback for the user. After getting the character’s response, users receive feedback through a dedicated panel in the chat interface. The feedback varies depending on the message option sent. If the message option sent is ambiguous or suggests the character's response was rude, the user receives constructive feedback. This is shown in the gray panel in the center of Figure \ref{fig:inc-feedback}. Constructive feedback is structured such that it first highlights the difference in interpretation/intent between the user and the character, identifies the most appropriate message option, and then explains why it is most appropriate. The user is also provided with a message that they can send to continue the conversation empathically, as shown at the bottom of Figure \ref{fig:inc-feedback}. If the user sends the most appropriate message option, positive feedback is provided to the user, as shown in the gray panel at the bottom of Figure \ref{fig:corr-feedback}. Positive feedback serves as encouragement, and explains why the other message options might lead to a misunderstanding.

\subsection{Development Process}
\label{sec:development-process}

\subsubsection{Advisory Board}
An advisory board of three autistic volunteers provided feedback on the design of NeuroBridge. All board members were enrolled as graduate and/or undergraduate students at Tufts University, and were identified through the authors’ existing network of autistic individuals who had previously participated in or expressed interest in serving on a board. Board members reviewed the prototype in three one-hour meetings held at the elementary, intermediate, and final stages of development. Each member evaluated NeuroBridge as a mock user and reviewed the AI-generated simulation, responses, and feedback, going through each simulated scenario at least twice. Each meeting was conducted by two researchers and recorded for later viewing. As detailed next, in-depth, open-ended feedback was obtained from the board and incorporated during development.

\subsubsection{Iterative Development}
Several improvements were made based on feedback from the advisory board. For instance, they recommended that when a user sends an unclear message, the AI character should ask a clarifying question like, `Do you mean X or Y?' rather than assuming one of X or Y, reflecting how they typically process uncertainty in real life. They also emphasized the importance of sharing these interpretations in more detail with neurotypical users in the feedback so that they can understand exactly how an autistic person might interpret language differently. They verified that two out of three message options in the simulated scenarios could, in fact, lead to a misunderstanding, while the remaining one was most appropriate. Additionally, the board reviewed the AI character's blunt responses and agreed that the tone reflected their past communication experiences, which sometimes led to negative reactions from others. Consistent with prior studies \cite{twips-haroon, milton-the}, they emphasized the importance of encouraging neurotypical individuals to understand autistic communication styles and viewed NeuroBridge's approach as effective.

Moreover, we also conducted five pilot studies with neurotypical users for preliminary testing and feedback. In the initial version of NeuroBridge, users had no control over the conversation topic or message composition; they were given a set topic and pre-determined message options to choose from. Based on feedback from pilot users, we added the ability for users to select a topic of interest and compose their own messages, which are then used to generate personalized scenarios, message options and feedback. Initially, we had also included filler turns, so that only every other message triggered a scenario, creating a more natural conversation flow. This nearly doubled the interaction time, so we eventually removed them. Additionally, we observed that our pilot users, including those knowledgeable about autistic communication, repeatedly re-read the message options and carefully reasoned through them, highlighting both the intended subtlety of the differences and the cognitive effort involved in making the correct selection.

\begin{figure*}[t]
    \centering

    \begin{subfigure}{\textwidth}
        \centering
         \fcolorbox{gray!60}{white}{\includegraphics[width=.9\textwidth]{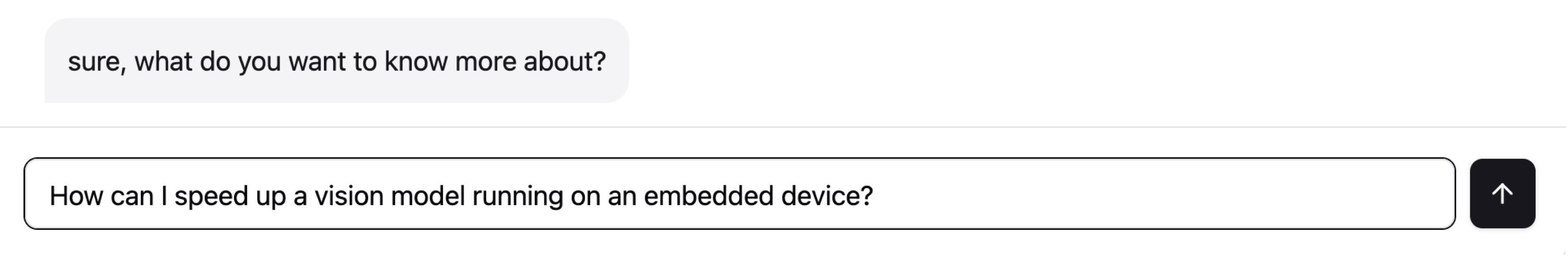}}
         \Description[Screenshot of the user input panel.]{Screenshot of the user input panel showing the AI character's latest message in a gray message bubble that reads, "sure, what do you want to know more about?" Below that, an input box shows the user has typed in, "How can I speed up a vision model running on an embedded device?" To the right of the input box is a send button.}
        \caption{Message crafted by user.}
        \label{fig:user-input}
    \end{subfigure}

    \vspace{1em}  

    \begin{subfigure}{\textwidth}
        \centering
         \fcolorbox{gray!60}{white}{\includegraphics[width=.9\textwidth]{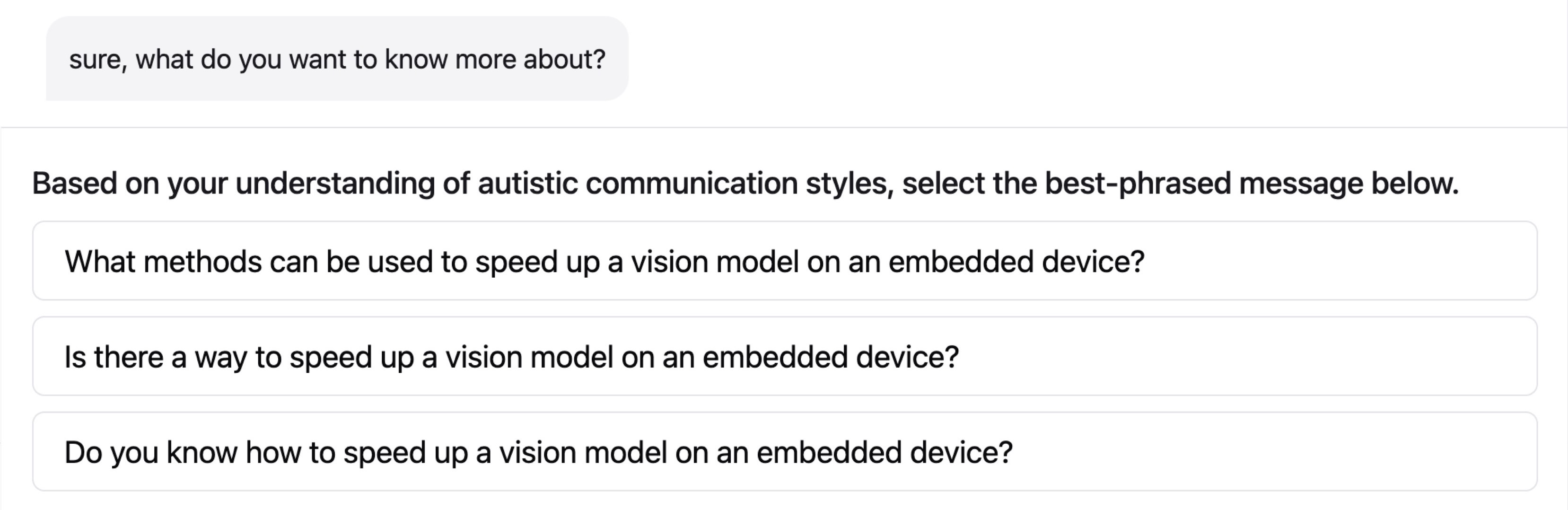}}
         \Description[Screenshot of the panel showing rephrased message options.]{Screenshot of the user input panel showing the AI character's latest message in a gray message bubble that reads, "sure, what do you want to know more about?" Below that, instructions to the user read, "Based on your understanding of autistic communication styles, select the best-phrased message below." Then, three options are presented: (a) "What methods can be used to speed up a vision model on an embedded device?" (b) "Is there a way to speed up a vision model on an embedded device?" and (c) "Do you know how to speed up a vision model on an embedded device?" }
        \caption{Rephrased message options based on message crafted by user.}
        \label{fig:rephrasings}
    \end{subfigure}

    \caption{The user is first prompted to input a message to send to the AI character, as shown in (a). Then, three unique variations are generated and displayed to the user, prompting them to select the best-phrased message, as shown in (b). }
    \label{fig:user-and-rephrasing}
\end{figure*}

\subsubsection{Prompting Strategy}
\label{sec:prompting-strategy}
LLMs take in input in the form of natural language, provided through `prompts'. A prompt is a carefully crafted instruction that guides the model's output. Following prior work, we iteratively refined the prompts for each task, such as generating message options, character responses and feedback \cite{goodman-lampost, twips-haroon}. Through repeated testing, we optimized them for consistency and reliability. Note that we did not instruct the LLM to act autistic or generate feedback from the perspective of an autistic person. Instead, we provided carefully crafted examples of message options, interpretations, and feedback for each scenario as `sample outputs' in the prompts. These examples helped ensure the LLM responded in the way we intended. Providing examples to improve output quality is an effective prompting technique and commonly known as few-shot learning \cite{parnami-learning}. We avoided referencing autism in any of the prompts to prevent perpetuating existing biases about autism in LLMs \cite{park-as}. All prompts have been made available as Supplementary Material for reproducibility.
\subsubsection{Implementation}

The frontend of NeuroBridge was developed using React and shadcn/ui, while the backend was built with FastAPI, incorporating both REST and WebSockets to facilitate real-time chat functionality. GPT-4o (GPT-4o-2024-0513 Regional) was used for LLM generation in all tasks, except for generating user message options involving emojis with variable interpretations. We used Claude 3.5 Sonnet (us.anthropic.claude-3-5-sonnet-20240620-v1:0) for it, as it outperformed GPT-4o on this task. Both models were accessed through a deployment on Microsoft Azure. The front-end was deployed on Cloudflare Pages, and the back-end was containerized using Docker and deployed on Google Cloud Run, with data storage managed through MongoDB Atlas.

\begin{figure*}[t]
    \centering
    \fcolorbox{gray!60}{white}{\includegraphics[width=.9\textwidth]{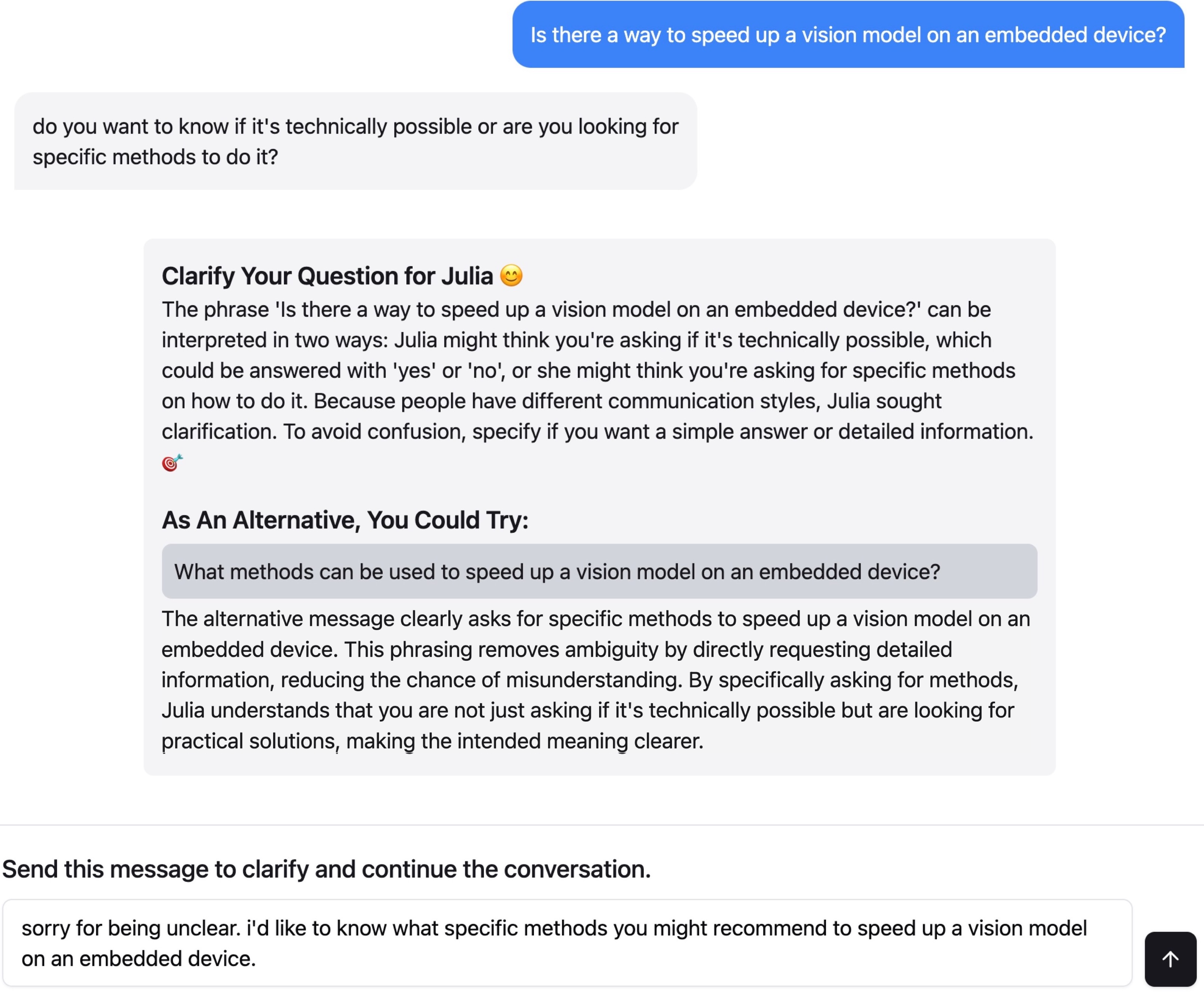}}
    \Description[Screenshot of the feedback after the user selects an option that is not the most appropriate choice.]{Screenshot of the feedback after the user selects an option that is not the most appropriate choice. A blue message bubble shows that the user sent, "Is there a way to speed up a vision model on an embedded device?" Under it, a gray message bubble shows that the AI character responded with "do you want to know if it's technically possible or are you looking for specific methods to do it?" Below, a feedback panel is displayed that starts with the heading "Clarify Your Question for Julia." This section reads, "The phrase 'Is there a way to speed up a vision model on an embedded device?' can be interpreted in two ways: Julia might think you're asking if it's technically possible, which could be answered with 'yes' or 'no', or she might think you're asking for specific methods on how to do it. Because people have different communication styles, Julia sought clarification. To avoid confusion, specify if you want a simple answer or detailed information." After this, a section titled "As An Alternative, You Could Try:" first shows an alternative message bubble, "What methods can be used to speed up a vision model on an embedded device?" Below this, it reads, "The alternative message clearly asks for specific methods to speed up a vision model on an embedded device. This phrasing removes ambiguity by directly requesting detailed information, reducing the chance of misunderstanding. By specifically asking for methods, Julia understands that Mark is not just asking if it's technically possible but is looking for practical solutions, making the intended meaning clearer." Under the feedback section, instructions to the user state, "Send this message to clarify and continue the conversation." An input box below this is filled in with, "sorry for being unclear. i'd like to know what specific methods you might recommend to speed up a vision model on an embedded device." To the right of this input box, there is a send button.}
    \caption{The message option sent by the user is shown in the blue message bubble. After, the gray message bubble shows that the AI character asks the user to clarify what they meant since an incorrect message option was sent. Then, the user receives a two-part constructive feedback (shown in the center gray panel) explaining why their choice was incorrect and why the other option was more appropriate. The user is then prompted to send the provided follow-up message to clarify and continue the conversation.}
    \label{fig:inc-feedback}
\end{figure*}

\section{Methodology}

In this section, we provide our positionality statement and an overview of the recruitment process, user study, and data collection and analysis methods. All study procedures were approved by the Institutional Review Board (IRB) at Tufts University.

\subsection{Positionality Statement}
We disclose that a majority of the authors identify as neurodivergent and have professional and/or personal relationships with neurodivergent people. Our approach is informed by the social model of disability, which frames disability as arising from societal structures rather than individual impairments. This perspective underpins our view that cross-neurotype communication breakdowns reflect differences in communication styles rather than deficits in social skills, shaping both the design and interpretation of this work.

\subsection{Recruitment}
Twelve neurotypical participants were recruited through flyers posted around Tufts University's main campus in Medford, MA, USA. Interested individuals completed a screening survey to determine eligibility. The inclusion criteria were: a) aged 18 or older, b) fluency in English (reading and writing), and c) ability to perform basic computer tasks. Participants were also asked about their familiarity with autistic communication styles in the screening survey, and an equal number were selected from each familiarity group. All participants identified as non-autistic. Participant information is shown in Table \ref{tab:participant-info}.

\subsection{User Study Overview}
User study sessions were conducted in person, on-campus in a lab setting, where participants were provided with a secure personal computer and monitor. Each session lasted about ninety minutes. Participants started by reading and signing a consent form describing the purpose and procedures of the study. Then, they proceeded to enter their name, pronouns, and a topic of interest, which was used to generate a social setting for the conversation with the AI character. After receiving instructions on how to use the interface, participants sent the first two messages to familiarize themselves with the system. No scenarios or feedback were triggered during this phase, as the first two were configured as test messages. This introductory phase allowed them to ask questions and get comfortable with the interface. Participants were encouraged to think aloud about their reasoning for selecting a message option, as well as their thoughts on the AI character’s responses and on the feedback they received in the remainder of the study. Participants interacted with the character until they had completed two rounds of each of the four scenarios.

\subsection{Data Collection and Analysis}

Participants' screen activity and audio were recorded during the user study. Upon completing the user study, participants took part in a semi-structured interview followed by a survey in the same session. The interviews were also audio-recorded. The 11-item Likert scale survey had statements rated on a 7-point scale from `Completely Disagree (1)' to `Completely Agree (7)'. The survey and interview delved into the usefulness of the simulation, its impact on participants' perceptions of autism, their attitudes toward AI feedback, and the effect of personalization on user engagement. The survey results provide an overview of self-reported user perceptions, while qualitative insights offer richer, explanatory insights about their experience using NeuroBridge. 

Given the small sample size (N=12), we report descriptive statistics (mean and standard deviation) for the survey results, along with verbatim survey statements and the cumulative percentage of responses indicating agreement (options ranging from 1 to 3, both inclusive) or disagreement (options ranging from 5 to 7, both inclusive) on the Likert scale in Figure \ref{fig:survey-results}. This approach is adapted from Goodman et al. and Adnin et al. \cite{goodman-lampost, adnin-i}. For qualitative analysis, we used Braun and Clarke’s thematic coding approach \cite{braun-using} with a deductive framework. Prior to the study, we developed the following set of deductive codes to categorize: perceptions of the simulation's usefulness; trust in the AI-generated simulation; reactions to AI-generated feedback; understanding and perceptions of autistic communication styles; and suggestions for improvement. A member of the research team first transcribed the audio data and then contextualized them with observations from the screen recordings. After importing the transcripts into NVivo \cite{dhakal-nvivo}, they extracted relevant quotes by reading the transcripts line by line, grouped them into themes, discussed the themes with other team members, and reviewed and refined them. Another member of the research team, who was not part of the initial study team, independently validated the themes and the data associated with each theme. A similar approach was used by Ahsen et al. and Haroon et al. \cite{ahsen-designing, twips-haroon}.

\section{Findings}
\label{sec:findings}
In this section, we discuss and synthesize our findings, supported by participants' quotations and relevant survey results.
\subsection{Usefulness of the Simulation Experience}

\subsubsection{Helps Develop Communication Awareness}


Several participants (P1, P3, P5, P6, P7, P8, P9, P11, P12) reported that taking part in the simulation helped them understand how an autistic person might interpret language differently. Many were surprised to see that these interpretations were plausible, and even obvious in hindsight, but never occurred to them. They highlighted that the AI character’s interpretation, included in the feedback, helped them understand exactly what part of their message could be received differently by an autistic person. For instance, P11 shared, \textit{"If the feedback just said `the figurative part in your message could cause confusion', I might’ve thought, `Okay, but why?' The example [interpretation] provided helps me understand what exactly Wendy [the AI character] is thinking when she is reading this."} Similarly, P3 reflected, \textit{"Explaining how the rocket emoji could be interpreted differently with an example [of an autistic interpretation] gave me a chance to see Jason’s [the AI Character] perspective."} Echoing these sentiments, P5 felt the feedback was useful for navigating future interactions. This was reflected in participants’ behavior as well. Upon encountering a similar scenario later in the simulation, most were able to identify the most appropriate response and referred to feedback from a previous turn to back their rationale. Overall, participants strongly agreed \textit{(avg. = 5.83, s.d = 1.19)} that the simulation helped them recognize the differences in communication styles of autistic and non-autistic individuals, as shown in row 1 of Figure \ref{fig:survey-results}.


\subsubsection{Closest to a Real Interaction with an Autistic Person}

 
Multiple participants (P5, P7, P9, P11) described the simulation as the closest they had come to interacting with an autistic person. They believed this was useful, as people often hold common misconceptions about autism that are unlikely to change without interacting with an autistic person in real-life. Since the simulation closely resembled such an experience, and because having open, exploratory conversations with an autistic person is not always possible, participants believed it served as an effective alternative. P7, who had in-depth knowledge of autistic communication through lived experience with their autistic sister, expressed \textit{"An interaction like this is probably the closest you can really get to emulating the experience of interacting with someone with autism."} They described the platform as a safe, low-stress environment for learning, and contrasted it with real-life interactions, \textit{"When interacting with someone with autism... things can kind of spiral out of control very quickly."} In contrast, \textit{"[With the simulation] you're sort of on some guardrails..."} Reflecting on their own experience, they added, \textit{"When I was growing up, this would have helped me a lot in understanding my autistic sister."} P5 echoed these sentiments, noting that the AI character’s responses allowed them to see how their message might have caused confusion for an autistic person if this was a real interaction, \textit{"You get to actually see what could happen if you say something that can cause confusion... it is very realistic, and prepares you to have a conversation [with an autistic person]."} Overall, participants agreed \textit{(avg. = 5.50, s.d = 1.45)} that the character's responses felt natural and realistic, as shown in row 11 of Figure \ref{fig:survey-results}.



\subsubsection{Enables Active Learning}

Several participants (P1, P2, P3, P4, P5, P6, P10, P12) appreciated the interactive nature of the platform, highlighting that it allowed them to apply what they were learning hands on. Having grown up with an autistic father, P6 had in-depth experience/knowledge of autistic communication styles. Yet, they expressed simply knowing wasn’t the same as applying that knowledge. The feedback they received on one of their responses revealed perspectives they hadn't considered, \textit{"I'm pretty knowledgeable on how autistic people communicate, but I didn't even think about how the chicken emoji could be interpreted like that [as described in the feedback]. After looking at the feedback, I was like, oh... yeah, you're right."} Similarly, P3 pointed out that conversing with the AI character exposed gaps in their knowledge, \textit{"It's not until you actually try to have a conversation that you can really see what they might not understand in what you say."} In addition, P1 noted that the process of crafting the response helped reinforce what they were learning, \textit{"I had to actually think about what the response [by the character] was, and how to best word my response to continue the conversation."} P3, P5, P6 and P10 echoed these sentiments. For P6, this was especially useful in moments of friction, \textit{"They [the character] said something in not the nicest tone... and I had to think through the response."} As shown in row 2 of Figure \ref{fig:survey-results}, participants generally agreed \textit{(avg. = 5.58, s.d = 1.83)} that the simulation would influence their future communication with autistic individuals.

\subsubsection{Personalized Feedback}

Several participants (P3, P5, P6, P9, P10, P11, P12) emphasized what made the simulation especially useful was the personalized nature of the feedback. Rather than presenting abstract or generic examples, the system provided feedback on messages that they had sent and were based on their original input. P11 explained, \textit{“What really helps is having the confusing parts of your own speech specifically pointed out.”} In addition, participants found value in seeing how the message they had come up with could be easily rephrased to cause or prevent a communication breakdown. In one instance, after reading the message options, P11 exclaimed, \textit{"A lot of the time when I was writing up my response, I didn’t even consider other ways to say the same thing. It is interesting to see what those options were and think about which of those made the most sense."} P12 echoed this sentiment, highlighting moments when the tool improved upon what they had tried to say, \textit{"I was okay with the way I worded myself, but it wasn’t perfect and then it would give me a better option that accomplished what I wanted to say in a very autistic-friendly way."} In this way, personalized feedback encouraged participants to reflect on their own communication style and assumptions.

\subsubsection{Engaging and Immersive}

Several participants (P1, P4, P7, P8, P9, P10, P11, P12) described the simulation as more engaging than other, common ways of gaining awareness, such as awareness blogs or videos. P9, who was already interested in chatbots, appreciated how the experience \textit{“replicates that feeling of talking to a real person”}, adding that it was \textit{“more engaging to have what feels like an actual conversation”} rather than passively consuming information. P11 echoed this sentiment, noting that reading felt \textit{“a lot more educational”}, whereas with the simulation, \textit{“you learn on the way.”} P4 described traditional formats as \textit{“passive”}, and P8 shared, \textit{“You read it, and then put it down and put it away, whereas this is a more memorable experience.”} Overall, as shown in row 3 of Figure \ref{fig:survey-results}, participants strongly preferred \textit{(avg. = 5.92, s.d = 1.51)} NeuroBridge over awareness blogs and videos. P12 further noted that reading or watching content can sometimes create a false sense of confidence, \textit{“It actually kind of harms you because you think, `Oh, I know what to avoid. I know what I need to do,’ and you don't realize that just knowing about it doesn't mean you actually know how to apply it.”}  They highlighted that in contrast, the simulation allows you to practice and test your understanding, making it easier to see what you truly grasp and where you might need to improve. For P1 and P7, the process of crafting their own responses kept them immersed throughout the simulation. As shown in rows 4 \textit{(negatively-worded, avg. = 1.50, s.d = 0.90)}, 5 \textit{(avg. = 6.92, s.d = 0.29)}, and 7 \textit{(avg. = 6.58, s.d = 0.67)} of Figure \ref{fig:survey-results}, users strongly valued the simulation's personalized and interactive format. In addition, participants generally did not feel the simulation was too time-consuming, as shown in row 6 \textit{(negatively-worded, avg. = 2.33, s.d = 1.23)} of Figure \ref{fig:survey-results}.

\begin{table*}[t]
\centering
\begin{tabularx}{\textwidth}{
  >{\hsize=0.5\hsize}X 
  >{\hsize=0.5\hsize}X 
  >{\hsize=1.2\hsize}X 
  >{\hsize=1.8\hsize}X
}
\toprule
\textbf{P\#} & \textbf{Age} & \textbf{Gender} & \textbf{Knowledge of Autistic Communication} \\
\midrule
\textbf{P1} & 18--24 & Female & I have no prior knowledge \\
\textbf{P2} & 18--24 & Female & I have heard of it but don't know much \\
\textbf{P3} & 18--24 & Female & I have heard of it but don't know much \\
\textbf{P4} & 18--24 & Female & I have a very basic understanding. \\
\textbf{P5} & 18--24 & Male & I have heard of it but don't know much \\
\textbf{P6} & 18--24 & Female & I have in-depth knowledge and/or experience \\
\textbf{P7} & 25--34 & Male & I have in-depth knowledge and/or experience \\
\textbf{P8} & 18--24 & Female & I have a very basic understanding. \\
\textbf{P9} & 18--24 & Female & I have in-depth knowledge and/or experience \\
\textbf{P10} & 18--24 & Male & I have no prior knowledge \\
\textbf{P11} & 18--24 & Non-binary/third gender & I have no prior knowledge \\
\textbf{P12} & 18--24 & Male & I have a very basic understanding. \\
\bottomrule
\end{tabularx}
\Description[A table shows the participant demographics and knowledge of autistic communication.]{A table shows the participant demographics and knowledge of autistic communication. All participants are aged 18-24, except one who is 25-34. Participants are mixed male and female, with one non-binary individual. There is also mixed prior knowledge of autistic communication.}
\vspace{0.5em}
\captionsetup{justification=raggedright,singlelinecheck=false}
\caption{Participant demographics and familiarity with autistic communication styles.}
\label{tab:participant-info}
\end{table*}
\subsection{Feelings of Trust and Skepticism in AI}

\subsubsection{Factors Affecting Trust in AI}

Several participants (P5, P6, P7, P9, P10) initially approached the simulation with skepticism because they were told it was powered by AI, but came to view it as trustworthy as they found their interaction with the AI character realistic, and the explanations provided by the system relatable and logically coherent. For instance, P5 described the AI character as \textit{"almost creepily realistic"}, and that it \textit{"easily could have been a real person."} They found the reasoning provided in the feedback convincing, particularly because it systematically explained how their message could be interpreted in different ways. The feedback was structured to first recognize the sender’s likely intent, then illustrate how and why the message might be received differently by the AI character, and finally suggest a better alternative along with a justification. This helped participants connect the dots between what they meant to say and how it could be misread. P5 reflected \textit{"It is the way it is explaining the phrases and the things I said... following a logical train of thought in its response."} P6 shared a similar view, stating that they found the feedback provided to them during the simulation trustworthy because it also aligned with their own reasoning.

Participants (P7, P8, P9, P11) also highlighted that their trust in the system was shaped by their personal background, such as their prior exposure to autistic individuals and technology. For example, P11, who had limited experience interacting with autistic people, shared that they trusted and were open to receiving constructive feedback from the system because they did not consider themselves knowledgeable on the topic. P8 noted that while they personally trusted the chatbot, their grandmother would likely be much more skeptical of it. In their view, prior experience with technology played a key role in whether someone would take the simulation seriously, \textit{“My 75-year-old grandma would probably be very skeptical of it... whereas if she was talking to a professor, or someone with autism, she would believe them without hesitation.”} Similarly, P7, an engineering student with an autistic sibling, admitted to having an \textit{"intrinsic"} bias against AI. They viewed it as a tool often misapplied to scientific problems beyond its limits, but found value in the simulation, \textit{“You kind of need to suspend disbelief. I know I’m talking to a machine, but it emulates it closely enough that I can get something out of it.”}  Their personal connection to autism allowed them to look past their skepticism of AI.

\begin{figure*}[!hbtp]
  \centering
  \includegraphics[width=.95\textwidth]{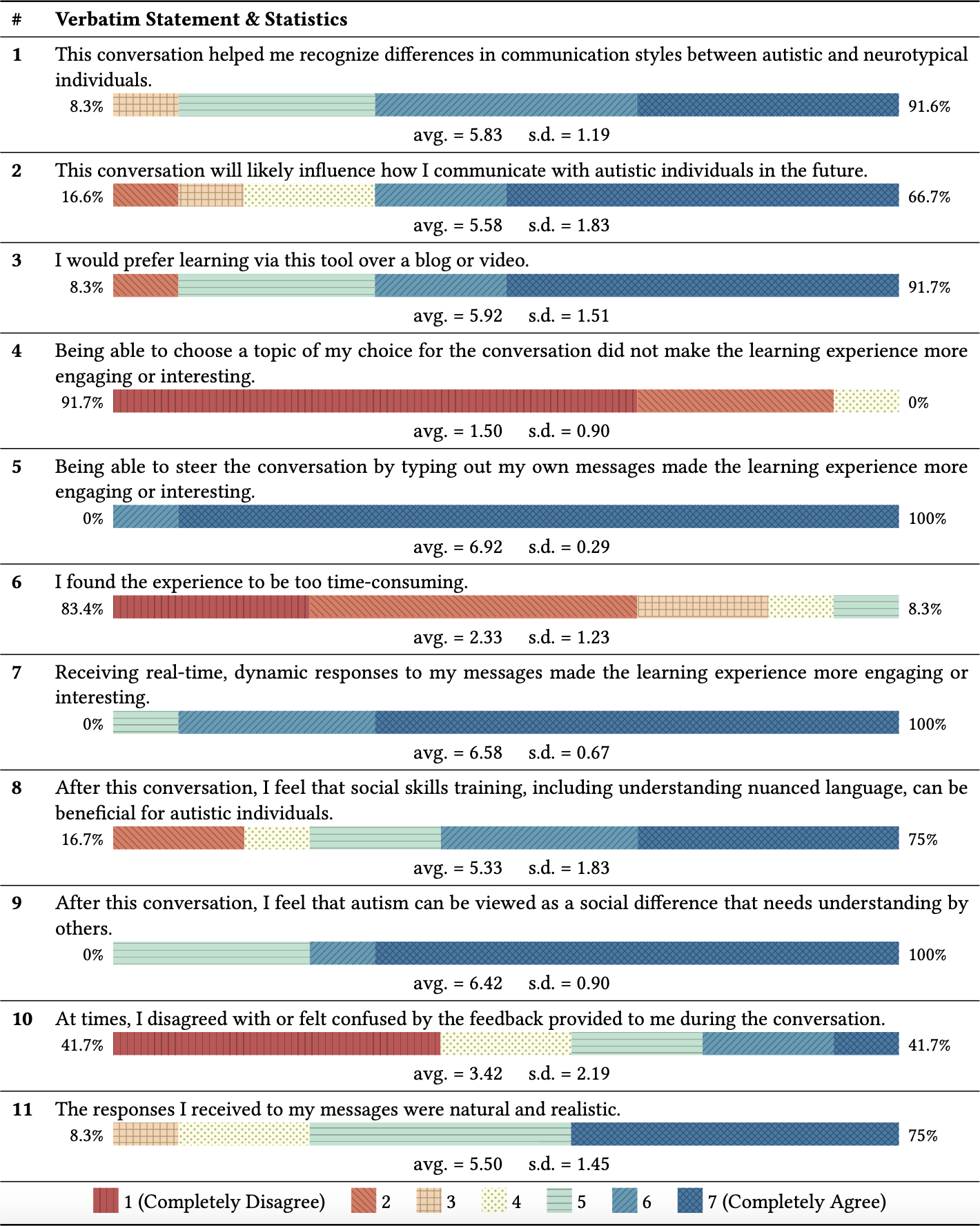}
    \Description[Participant responses to the survey.]{Participant responses to statements 1 through 11 from the survey are summarized using descriptive statistics in a table. For each statement, the average and standard deviations are reported, along with the percentage of respondents who selected values between 1 and 3, and those who selected values between 4 and 7.}
    \caption{Survey results with verbatim statements and statistics. The percentage on the left represents the number of participants who selected values between 1 and 3 (both inclusive), while the percentage on the right represents the number of participants who selected values between 5 and 7 (both inclusive). Responses of 4 (middle) are excluded from both percentages.}
  \label{fig:survey-results}
\end{figure*}
\subsubsection{Reactions to AI Feedback}

At most occasions, participants described feeling curious, open, and motivated when they received AI-generated feedback. For instance, P12 reflected, \textit{“It makes me curious, like, how can I, going back into real life, interacting with actual autistic people, tailor my language to make sure I'm communicating with them effectively?”} In particular, participants appreciated that communication differences were framed constructively in the feedback, without labeling their response as \textit{“wrong”}. P3 echoed this sentiment and noted that even when they did not perform well, the system recognized that they were trying, \textit{“Even when I say something wrong, it isn't like, `You're wrong.’ Even the titles are `Thoughtful Communication’ and `A Small Tweak to Make Your Message Clear’. They're very much acknowledging that you are trying.”} The use of emojis and a supportive tone contributed to making the feedback feel friendly and supportive, helping participants stay open and receptive. As P2 expressed, \textit{“I like the little star emoji [in the feedback]. It adds a nice little bit of flair and makes it feel like a little more celebratory."} This sentiment was echoed by P5 and P9.

In addition, participants found it useful to receive feedback not only when they failed to identify the most appropriate message option, but also when they succeeded. P12 elaborated, \textit{"I definitely think you should continue to provide feedback when things are going well. I get so frustrated when I only get feedback for doing something wrong. I want to know what I did well so I can keep doing it in the future. I want to know exactly what part of my behavior was good, not just `your behavior is good, keep doing it', because otherwise, I’m not really sure what to continue."} Participants noted that positive feedback was not only encouraging, but also helpful for learning. This was particularly important for users unfamiliar with autism. While they might have selected the correct answer, they could have done so without fully understanding why. The feedback helped validate their reasoning and fill in any gaps in understanding. P4 reflected, \textit{“If I don’t know much about autistic communication, I might pick the right option for the wrong reason. So it’s helpful to hear, `Yes, this is right and here’s why.’”}

However, on a few occasions, participants expressed feeling defensive, describing the feedback as instructive and diminishing their sense of agency. For example, P10 remarked, \textit{“The phrasing of the feedback should come off a bit more neutral. Some lines come off as almost an attack on how you talk, especially when some people... may go into this with no prior experience interacting with someone with autism.”} P7, who had lived experience supporting an autistic family member, expressed that frustration and defensiveness are natural in cross-neurotype communication. They emphasized the need for the feedback to not only offer constructive suggestions, but also to validate these emotions, \textit{“For me, a big part of it is validating those feelings... You should insert something like, `It is okay to feel frustrated sometimes, you are human too’... and then go into, `Here’s how you can be better and kind of meet them halfway.’”} Similarly, P12 described one instance where they felt sidelined in the interaction, \textit{“It [the feedback] really frustrates me because I feel like it puts too much focus on Autumn [the AI character] and takes agency away from me... It feels like you're just playing to Autumn's whims.”} Notably, all of these reactions were observed during the scenario around misperceived bluntness.

\subsubsection{Cannot Substitute Real Interactions}
Participants (P7, P8, P11) emphasized that while the AI-driven simulation was useful, it was still important to hear directly from autistic individuals, rather than relying solely on an AI to represent them. P8, who had limited personal experience with autistic people, felt that the chatbot helped illustrate key communication pitfalls and did a good job of showing how seemingly clear messages could be received differently, but ultimately concluded, \textit{“as a whole, having an experience with a person is a better way for getting to know them.”} P7 echoed this sentiment, framing the tool as part of a larger learning journey, \textit{“If you wanted to create a package of how to interact with autistic people one-on-one, this would be an element of that, but it wouldn’t be the whole thing.”} They appreciated the simulation’s ability to model scenarios and spark reflection, but felt it could only approximate the complex dynamics involved in a real conversation.

\subsection{Concerns and Improvements}

\subsubsection{Perceptions of Autism}
We were particularly interested in how the simulation shaped participants’ perceptions of autistic abilities. Survey results show that participants strongly agreed \textit {(avg. = 6.42, s.d = 0.90)} that "autism can be viewed as a social difference that needs understanding by others" after the simulation, as shown in row 9 of Figure \ref{fig:survey-results}. However, participants also expressed agreement \textit {(avg. = 5.33, s.d = 1.83)} with the statement, "social skills training, including understanding nuanced language, can be beneficial for autistic individuals", as shown in row 8 of Figure \ref{fig:survey-results}. Qualitative results help contextualize this; some of our participants came away with reinforced stereotypes about autism. For example, P10 remarked that the AI character’s responses made them feel its text comprehension abilities as \textit{“a bit below average”}, especially when it took common metaphors too literally. Similarly, P9 expressed concern that some users might interpret this behavior as a sign of cognitive inferiority, and stressed these literal interpretations need to be framed as a difference (as opposed to a deficiency) more concretely in the feedback. Similarly, P2 and P6 wondered whether the AI was underestimating autistic people's abilities related to symbolic understanding, as they felt emojis like a thumbs-up or fire icon didn’t seem inherently complex, yet were treated as such by the AI character. In contrast, P9 agreed that while these emojis could be confusing depending on the context in which they are used, they acknowledged the risk that users unfamiliar with autism might misread these incidents as evidence of limited ability.

During our meetings with the advisory board, we had reviewed several examples that neurotypical participants found to be too simple to be misunderstood, such as one involving a basic emoji. Members of the board pointed out that things that appear simple on the surface can be confusing depending on the context in which they are used. This reveals how neurotypical individuals may struggle to recognize that expressions they consider straightforward can be confusing for autistic individuals. Nonetheless, P9 made an interesting observation; although the AI character was configured to be literal, it did end up using metaphors once or twice. P9 felt this challenged the assumption that autistic individuals have below-average language skills or cannot understand/use figurative language, because the character was shown using it a few times. In P9’s view, this prevented a stereotypical portrayal of autism from being reinforced, while still highlighting that figurative language may not always be the preferred option.

\subsubsection{Generating Message Options.}One of the LLM's core tasks was to generate alternative versions of the user's message that were semantically identical but phrased differently, depending on the given scenario. However, several participants raised concerns about the quality of the message options, particularly in the scenario related to emojis with variable interpretations. Participants (P1, P5, P6, P7, P8, P12) found that the emojis added by the LLM often felt random or disconnected from the content of the message. For example, P12 described the use of crystal ball and alien emojis as \textit{“super, super, weird”}, stating they wouldn’t have understood the purpose of adding them without reading the explanations in the feedback. Similarly, P1 stated that some of the emojis \textit{"felt out of place"} and would confuse neurotypical individuals as much as autistic people. P7 expressed frustration at being \textit{"forced into a series of bad options"}, highlighting a mismatch between the emoji’s tone and the content of the messages. 
Participants acknowledged that eventually the feedback helped clarify why those emojis were added, but the feedback was revealed to them only after they had sent the message. P8 wondered whether such abstract associations would be ever be apparent to anyone without the feedback. Overall, nearly 40\% of participants expressed some degree of confusion \textit{(negatively-worded, avg. = 3.42, s.d = 2.19)} during the simulation, as shown in row 10 of Figure \ref{fig:survey-results}.

\subsubsection{Modeling the Blunt Scenario.} Another key task for the LLM was to craft a blunt message on behalf of the character that would serve as a turning point in the conversation. This message was intended to simulate a situation in which the character might be perceived as blunt by the participant, triggering a harsh or confrontational response from them. However, several participants (P1, P2, P8, P11, P12) stated that these trigger messages did not always come off as blunt. Participants described the tone of these messages as \textit{“neutral”}, \textit{“factual”} or \textit{“reasonable”} depending on the context. P12, for example, stated, \textit{“They do not seem to me to be blunt... it’s a simple statement. They’re not elaborating, but they’re also completely answering my question.”} P1 similarly downplayed any negative tone, saying, \textit{“I wouldn’t have thought that he [the AI character] was being blunt, or, you know… rude in any way.”} P2 added that such directness felt familiar and unremarkable, \textit{“I’m used to hearing people say things like that... it seems neutral. It seems factual.”} As a result, some participants were confused about why they were presented with confrontational message options. P8, for instance, felt that message options like `What’s with the attitude?' did not align with their interpretation of the AI character's message. \textit{"Those surprised me as being options,"} they explained, \textit{"because I didn’t interpret that [the trigger message] at all as giving attitude or being dismissive in any way."} P11 described a moment where the AI character seemed to contradict itself by first saying, \textit{"Do you want to hear about my experiences? I think they're interesting,"} and then suddenly following with, \textit{"They’re not interesting. Why do you want to know?"} The inconsistency left P11 confused, \textit{"It’s like almost contradicting the text they just sent."} In this instance, the LLM struggled to maintain conversational flow and logical coherence


\section{Discussion}

In this section, we reflect on our findings and discuss design implications for representing disabilities through AI, the need for making NeuroBridge more personalized, and LLMs' limitations in modeling complex social scenarios.
\subsection{Representing Disabilities through AI}

A key challenge in creating accurate and complete representations of disability through AI lies in capturing the diversity of disabled people's lived experiences \cite{tigwell-nuanced}. This is especially true for autism, which spans a broad spectrum characterized by nuanced, and often subtle, differences. While the scenarios in NeuroBridge reflect common challenges experienced by autistic individuals with a direct and literal communication style, not everyone with this style finds them difficult to navigate. Moreover, other scenarios, such as those involving sarcasm or sexual innuendos, could be incorporated too \cite{figurative-lampri}. Our participants were observant of this limitation, and suggested incorporating multiple AI characters to represent a broader range of scenarios and communication styles -- echoing prior work that highlights how single-perspective disability representations can unintentionally reinforce stereotypes \cite{tigwell-nuanced} and misconceptions \cite{kiger-disability}. Additionally, participants recommended adding in-situ `citations' with each scenario, such as links to Reddit threads or first-hand accounts from autistic individuals. This would not only enhance the credibility, transparency, and grounding of the AI-generated simulation, but also expose users to everyday experiences beyond those represented in the simulation. Understanding this context can help neurotypical individuals better understand how disabled individuals truly feel and identify with their disability \cite{tigwell-nuanced, barney-disability}. While gaps remain and it may be difficult to capture every nuance, LLMs’ ability to simulate diverse communication styles is a meaningful step toward more accurate disability representations in AI.

\subsection{The Need for Situational Context}

Our findings highlight the importance of LLM-powered interactivity, personalization, and realism in sustaining user engagement and active learning \cite{ha-clochat}. Informed by these insights, we propose incorporating `situational context' into the simulation by situating communication scenarios within specific social roles or relationships \cite{burgoon-expectancy}, such as student-teacher or doctor-patient dyads. Note that appropriateness varies based on social dynamics; bluntness may be acceptable among friends but is generally less so between a teacher and a student. Incorporating situational context helps capture these nuances more accurately, while also raising the question of how individuals in authority roles, such as teachers or doctors, respond to feedback provided to them during the simulation. Our findings suggest factors like familiarity with autism and/or AI can affect user experience and attitude. Hence, it will be useful to examine how authority influences openness to critical feedback. Moreover, incorporating situational context could enhance transfer of knowledge and awareness from the simulation to real-world interactions. For example, simulating a disagreement between a student (role-played by AI) and teacher (role-played by a human), could be particularly beneficial for a real-life teacher, as they may face a similar situation at work \cite{levi-keren-simulation}. Prior work shows that autistic individuals often use AI tools in hierarchical settings, where the risks and consequences of miscommunication are amplified \cite{jang-its}. Training neurotypical users in these scenarios will also help them recognize the higher stakes involved and the need for thoughtful, considerate communication.

\subsection{The Fine Line in Trusting AI}

Participants readily placed their trust in the AI-generated simulation and feedback, despite initially approaching it with skepticism. This was particularly observed among individuals with limited prior knowledge of autism. Given that LLMs have been shown to perpetuate biases against disabled individuals, including those on the autism spectrum \cite{gadiraju-i, park-as}, it is crucial for users to calibrate \cite{liao-designing} the amount of trust they place in AI-generated representations of disabilities. Some of our participants suggested the simulation should be paired with preparatory materials, such as a primer on autism, so that they feel more confident going into it and can view it from a critical lens. In addition, future iterations of NeuroBridge could consider incorporating features to facilitate structured and systematic reflection. These could include online discussion or chat features for engaging with other users or autistic/expert moderators. While LLM biases related to autism detection and demographics have been explored \cite{park-as}, exploring how LLMs simulate autistic communication styles with minimal prompting (as opposed to our approach, which involved extensive instruction and no explicit mention of autism) warrants further investigation and could uncover additional biases.

\subsection{Challenges of AI-driven Simulations}
The LLM performed well for most tasks, but when failures occurred, they were often due to dependencies across tasks even though task decomposition has been shown to improve LLM performance \cite{khot-decomposed}. For example, if the AI-generated message options (the first task) did not reflect the nuances of the scenario to be simulated very well, the LLM struggled to later provide a convincing explanation (the second task) for why those options might be perceived as confusing. This resulted in a trickle-down effect, with issues in the early stages undermining performance in later stages. For us, this posed a challenge as multiple components of NeuroBridge rely on each other to coherently scaffold the simulation. Interestingly, since much of the simulation's content was open to interpretation, users often formed their own conclusions and were somewhat open to the AI's different or even incorrect interpretations, thinking they might be valid as well. This observation aligns with prior work suggesting users may overly ascribe intent to AI, a phenomenon known as `algorithmic anthropomorphism' \cite{cohn-believing}. In some cases, participants' perceptions of autism were negatively influenced by their perceptions of the LLM's capabilities \cite{kiger-disability}. For example, a few participants speculated that the AI had malfunctioned when they encountered a scenario they felt was too simple to be misunderstood by anyone. In this way, how users perceive AI may directly impact how they view the identities it represents.

\subsection{Limitations}

There are a number of limitations of our study. First, recruiting participants from a university setting limits the generalizability of our findings, as individuals from diverse age groups, backgrounds, and education levels may be less open to change, and as a result, react differently to feedback provided to them in the simulation. Hence, while our analysis reveals recurring themes, a broader demographic could uncover additional themes. Moreover, the study relied primarily on self-reported data, which may introduce bias as participants may not fully disclose their opinions. Incorporating more objective pre-post measures could offer deeper insights into the intervention’s effectiveness. Future research should also examine its long-term impact by exploring how it influences neurotypical individuals’ behavior in real-world interactions, particularly among family and friends of autistic people. Finally, we were only able to incorporate a limited set of communication scenarios, and a more comprehensive implementation would include a wider range.

\section{Conclusion}
In this paper, we argue for redressing the disproportionate burden autistic people bear in adapting to neurotypical communication norms. To this end, we present the design and evaluation of NeuroBridge, an LLM-powered tool that helps neurotypical individuals better understand autistic forms of expression, and reflect on their role in shaping cross-neurotype interactions. In a user study with 12 neurotypical participants, we find that NeuroBridge improved their understanding of how autistic people may interpret language differently, with all describing autism as a social difference that “needs understanding by others” after completing the simulation. Participants valued the simulation’s personalized, interactive format, and even those familiar with autism through lived experience reported gaining new knowledge about autistic communication. While many found AI-generated feedback to be "constructive", "logical", and "non-judgmental", participants also reported feeling defensive on a few occasions. Most perceived the portrayal of autism in the simulation as accurate, indicating that users may readily accept AI-generated (mis)representations of disabilities. Despite strong overall performance, our results show that LLMs are more adept at simulating certain social scenarios than others. We conclude by discussing design implications for disability representation in AI, the need for making NeuroBridge more personalized, and the limitations of LLMs in modeling complex social interactions.

\section{Acknowledgments}
We thank Ayesha Naeem, our advisory board, all members of the NAT lab and Crehan lab, and the anonymous ASSETS reviewers for their invaluable feedback that helped improve this work. This work was partially supported by a Tufts Springboard award.

\bibliographystyle{unsrt}
\bibliography{bibliography}
\pagebreak
\onecolumn
\appendix

\section{Simulation Flows}
\label{appendix:simulation-flows}
Interaction flows for the figurative expression, emoji with variable interpretations, and being misperceived as blunt scenarios are presented below. The interaction flow for the indirect speech acts scenario is discussed in Section \ref{sec:development-process} and Figures \ref{fig:user-and-rephrasing} and \ref{fig:inc-feedback}.

\subsection{Figurative Expression}

\begin{figure*}[h]
    \centering
    \includegraphics[width=.9\textwidth]{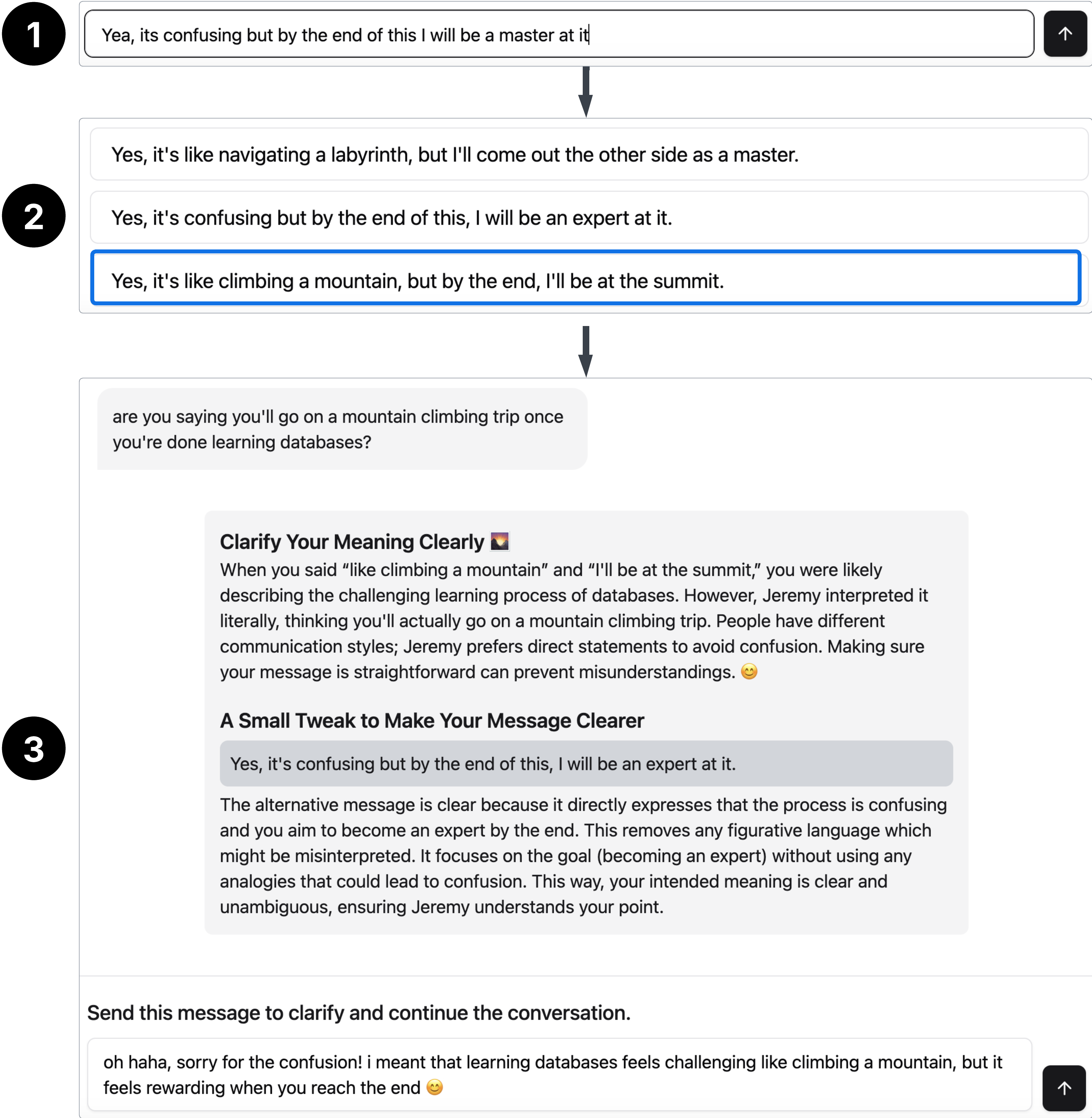}
    \Description[Screenshot shows the flow from user input to feedback.]{Screenshot shows (1) "Yea, it's confusing but by the end of this, I will be a master at it" is inputted by the user into a text box. (2) three options are shown: (a) "Yes, it's like navigating a labyrinth, but I'll come out the other side as a master."; (b) "Yes, it's confusing but by the end of this, I will be an expert at it."; and (c) "Yes, it's like climbing a mountain, but by the end, I'll be at the summit." Option (c) is selected. (3) The AI character responds with "are you saying you'll go on a mountain climbing trip once you're done learning databases?" Then, there is a feedback section titled "Clarify Your Meaning Clearly." Under this, it says, "When you said "like climbing a mountain" and "I'll be at the summit," you were likely describing the challenging learning process of databases. However, Jeremy interpreted it literally, thinking you'll actually go on a mountain climbing trip. People have different communication styles; Jeremy prefers direct statements to avoid confusion. Making sure your message is straightforward can prevent misunderstandings." Under this, there is a heading "A Small Tweak to Make Your Message Clearer." Then, a box that shows the alternative message, "Sure, what would you like to discuss?" Under this, a paragraph states, "The alternative message is clear because it directly expresses that the process is confusing and you aim to become an expert by the end. This removes any figurative language which might be misinterpreted. It focuses on the goal (becoming an expert) without using any analogies that could lead to confusion. This way, your intended meaning is clear and unambiguous, ensuring Jeremy understands your point." Under the feedback panel, a user input panel shows, "Send this message to clarify and continue the conversation." The input box is filled with "oh haha, sorry for the confusion! i meant that learning databases feels challenging like climbing a mountain, but it feels rewarding when you reach the end"}
    \caption{The simulation flow when an incorrect message option is sent in the "figurative expression" scenario. (1) shows the original message the user typed in; (2) shows the three message options generated and the user's choice; and (3) shows the AI character's response to that message and the feedback received.}
    \label{fig:incorrect-example-figurative}
\end{figure*}

\pagebreak
\subsection{Emoji with Variable Interpretation}

\begin{figure*}[h]
    \centering
    \includegraphics[width=.9\textwidth]{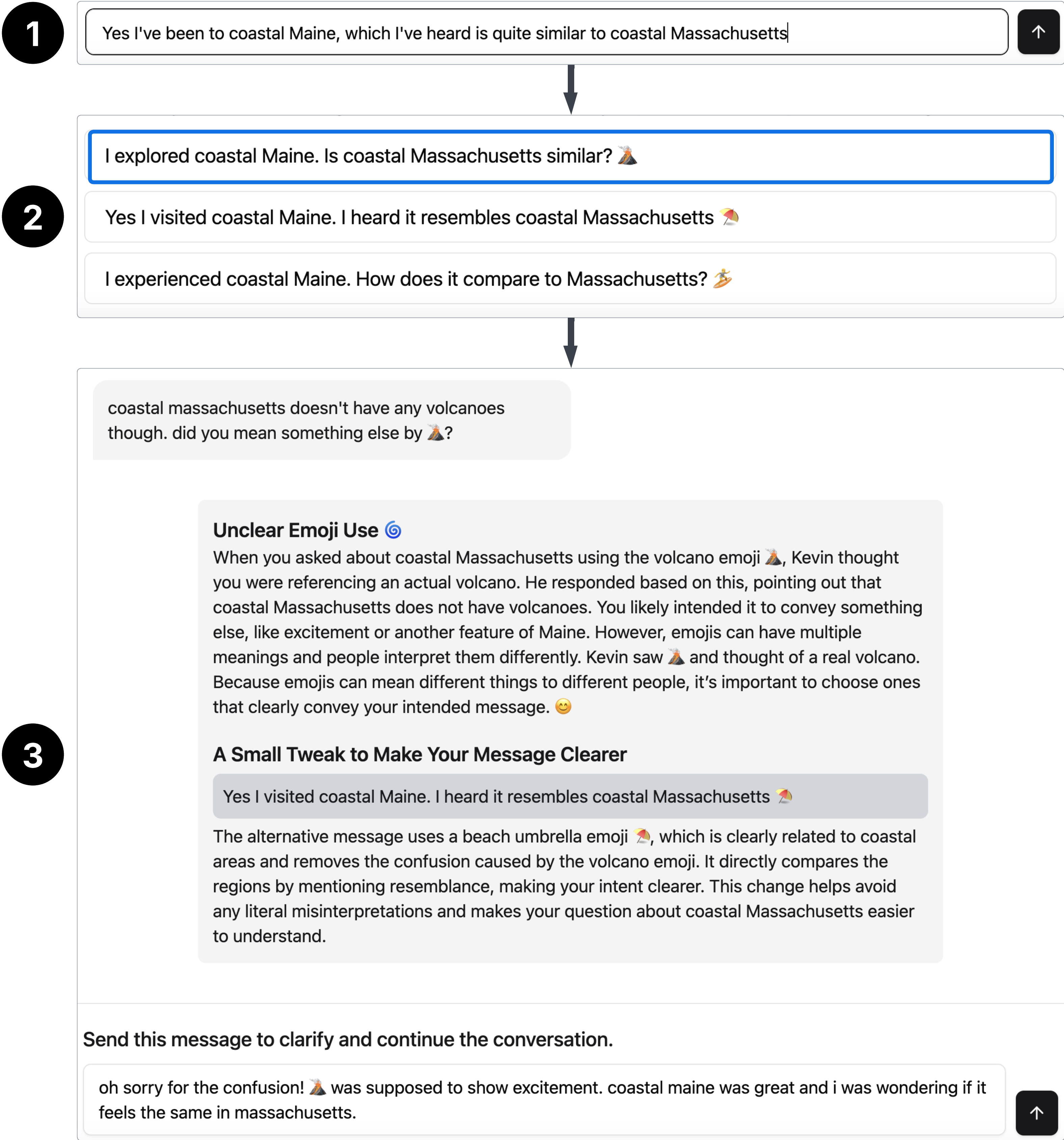}
    \Description[Screenshot shows the flow from user input to feedback.]{Screenshot shows (1) "Yes, I've been to coastal Maine, which I've heard is quite similar to coastal Massachusetts" is inputted by the user into a text box. (2) three options are shown: (a) "I explored coastal Maine. Is coastal Massachusetts similar? [volcano emoji]"; (b) "Yes I visited coastal Maine. I heard it resembles coastal Massachusetts [beach umbrella emoji]"; and (c) "I experienced coastal Maine. How does it compare to Massachusetts? [person surfing emoji]" Option (a) is selected. (3) The AI character responds with "coastal massachusetts doesn't have any volcanoes though. did you mean something else by [volcano emoji]?" Then, there is a feedback section titled "Unclear Emoji Use." Under this, there is a heading "A Small Tweak to Make Your Message Clearer." Then, a box that shows the alternative message, "Yes I visited coastal Maine. I heard it resembles coastal Massachusetts [beach umbrella emoji]" Under this, a paragraph states, "The alternative message uses a beach umbrella emoji, which is clearly related to coastal areas and removes the confusion caused by the volcano emoji. It directly compares the regions by mentioning resemblance, making your intent clearer. This change helps avoid any literal misinterpretations and makes your question about coastal Massachusetts easier to understand." Under the feedback panel, a user input panel shows, "Send this message to clarify and continue the conversation." The input box is filled with "oh sorry for the confusion! [volcano emoji] was supposed to show excitement. coastal maine was great and i was wondering if it feels the same in massachusetts."}
    \caption{The simulation flow when an incorrect message option is sent in the "emoji with variable interpretations" scenario. (1) shows the original message the user typed in; (2) shows the three message options generated and the user's choice; and (3) shows the AI character's response to that message and the feedback received.}
    \label{fig:incorrect-example-emoji}
\end{figure*}

\pagebreak
\subsection{Misperceived As Blunt}

\begin{figure*}[h]
    \centering
    \includegraphics[width=.9\textwidth]{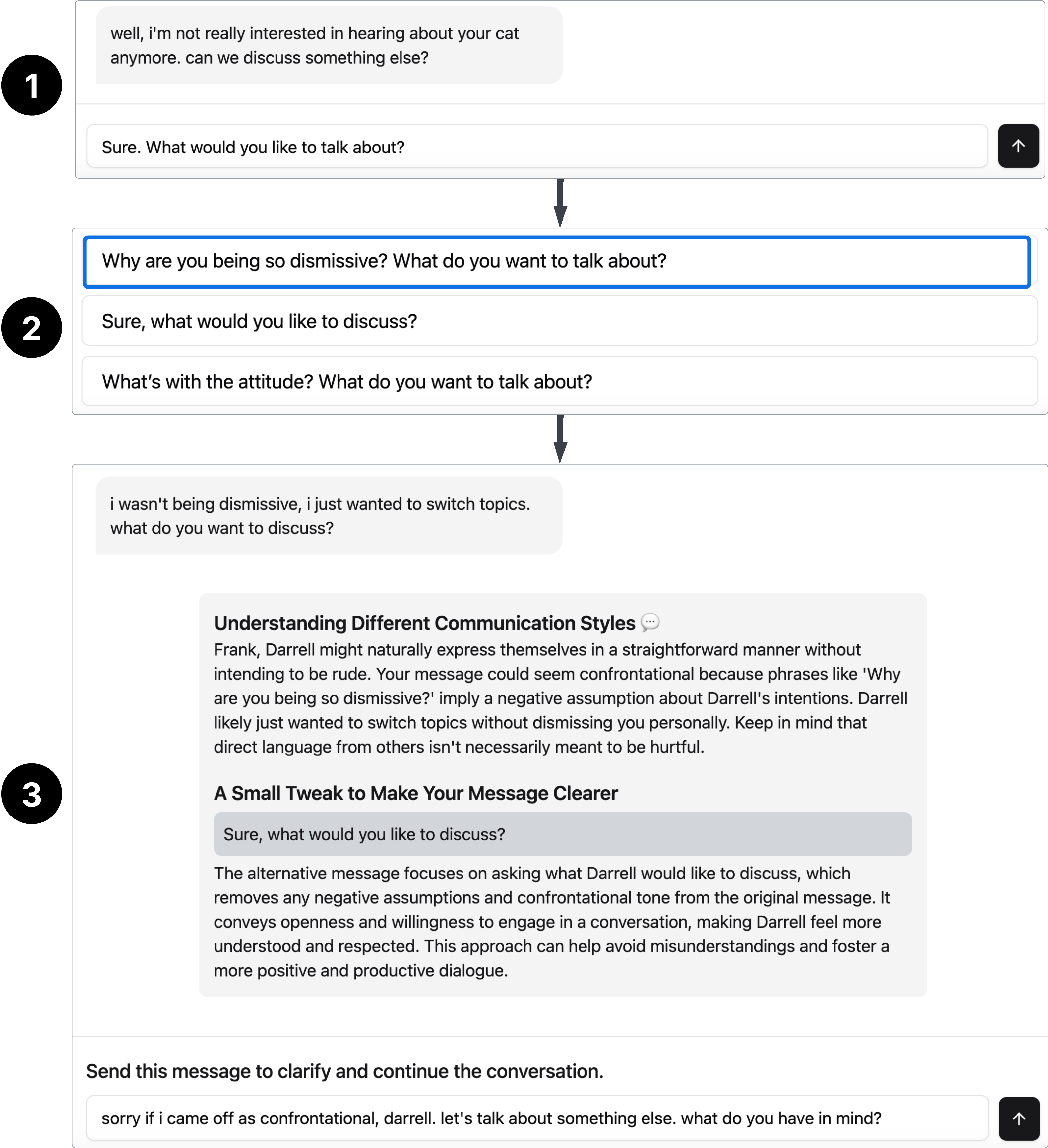}
    \Description[Screenshot shows the flow from user input to feedback.]{Screenshot shows (1) The AI character's latest message, "well, i'm not really interested in hearing about your cat anymore. can we discuss something else?" is shown in a gray message bubble. In a user input panel, "Sure. What would you like to talk about?" is inputted by the user into a text box. (2) three options are shown: (a) "Why are you being so dismissive? What do you want to talk about?"; (b) "Sure, what would you like to discuss?"; and (c) "What's with the attitude? What do you want to talk about?" Option (a) is selected. (3) The AI character responds with "i wasn't being dismissive, i just wanted to switch topics. what do you want to discuss?" Then, there is a feedback section titled "Understanding Different Communication Styles." Under this, it says, "Frank, Darrell might naturally express themselves in a straightforward manner without intending to be rude. Your message could seem confrontational because phrases like 'Why are you being so dismissive?' imply a negative assumption about Darrell's intentions. Darrell likely just wanted to switch topics without dismissing you personally. Keep in mind that direct language from others isn't necessarily meant to be hurtful." Under this, there is a heading "A Small Tweak to Make Your Message Clearer." Then, a box that shows the alternative message, "Sure, what would you like to discuss?" Under this, there is a heading "A Small Tweak to Make Your Message Clearer." Then, a box that shows the alternative message, "Yes I visited coastal Maine. I heard it resembles coastal Massachusetts [beach umbrella emoji]" Under this, a paragraph states, "The alternative message focuses on asking what Darrell would like to discuss, which removes any negative assumptions and confrontational tone from the original message. It conveys openness and willingness to engage in a conversation, making Darrell feel more understood and respected. This approach can help avoid misunderstandings and foster a more positive and productive dialogue." Under the feedback panel, a user input panel shows, "Send this message to clarify and continue the conversation." The input box is filled with "sorry if i came off as confrontational, darrell. let's talk about something else. what do you have in mind?"}
    \caption{The simulation flow when an incorrect message option is sent in the "misperceived as blunt scenario". (1) shows the AI character's blunt message and the original message the user typed in; (2) shows the three message options generated and the user's choice; and (3) shows the AI character's response to that message and the feedback received.}
    \label{fig:incorrect-example-blunt}
\end{figure*}

\end{document}